\documentclass[sigconf, screen]{acmart}

\usepackage[]{hyperref}
\usepackage[]{hyperxmp}
\usepackage{algorithm}
\usepackage{algorithmic}
\usepackage{url}

\AtBeginDocument{%
  }

\setcopyright{none} 
\acmISBN{}
\acmConference [Huawei Technologies Co]{}{.Ltd.}

\begin{document}

\title{DEER: Deep Runahead for Instruction Prefetching on Modern Mobile Workloads}

\author{Parmida Vahdatniya}
\affiliation{
\institution{Huawei Technologies Co,.Ltd.}
\country{Canada}
}
\email{parmida.829667@huawei.com}

\author{Julian Humecki}
\affiliation{
\institution{Huawei Technologies Co,.Ltd.}
\country{Canada}
}
\email{julian.humecki1@h-partners.com}

\author{Henry Kao}
\affiliation{
\institution{Huawei Technologies Co,.Ltd.}
\country{Canada}
}
\email{henry.kao1@huawei.com}

\author{Tony Li}
\affiliation{
\institution{Huawei Technologies Co,.Ltd.}
\country{Canada}
}
\email{tony.li2@huawei.com}

\author{Ali Sedaghati}
\affiliation{
\institution{Huawei Technologies Co,.Ltd.}
\country{Canada}
}
\email{ali.829657@huawei.com}

\author{Fang Su}
\affiliation{
\institution{Huawei Technologies Co,.Ltd.}
\country{China}
}
\email{fang.su@huawei.com}

\author{Ruoyu Zhou}
\affiliation{
\institution{Huawei Technologies Co,.Ltd.}
\country{China}
}
\email{ruoyu.zhou@hisilicon.com}

\author{Alex Bi}
\affiliation{
\institution{Huawei Technologies Co,.Ltd.}
\country{China}
}
\email{alex.bi@huawei.com}

\author{Reza Azimi}
\affiliation{
\institution{Huawei Technologies Co,.Ltd.}
\country{Canada}
}
\email{reza.azimi1@huawei.com}

\author{Maziar Goudarzi}
\affiliation{
\institution{Huawei Technologies Co,.Ltd.}
\country{Canada}
}
\email{maziar.goudarzi@huawei.com}

\begin{abstract}
Mobile workloads incur heavy frontend stalls due to increasingly large code footprints as well as long repeat cycles. 
Existing instruction-prefetching techniques suffer from low coverage, poor timeliness, or high cost.
We provide a SW/HW co-designed I-prefetcher; DEER uses profile analysis to extract metadata information that allow the hardware to prefetch the most likely future instruction cachelines, hundreds of instructions earlier. This profile analysis skips over loops and recursions to go deeper into the future, and uses a return-address stack on the hardware side to allow prefetch on the return-path from large call-stacks. 
The produced metadata table is put in DRAM, pointed to by an in-hardware register; 
the high depth of the lookahead allows to preload the metadata in time and thus nearly no on-chip metadata storage is needed.
Gem5 evaluation on real-world modern mobile workloads shows up to 45\% reduction in L2 instruction-miss rate (19.6\% on average), resulting in up to 8\% speedup (4.7\% on average). These gains are up to 4X larger than full-hardware record-and-replay prefetchers, while needing two orders of magnitude smaller on-chip storage.
\end{abstract}

\maketitle

\section{Introduction}
\label{intro}

Code footprint is growing faster than hardware tables can be economically enlarged to cope with~\cite{asmdb, ayers2018memory}, making frontend stalls a critical performance bottleneck  ~\cite{kanev2015profiling,ferdman2012clearing,schall2022lukewarm,luk1998cooperative}.
Modern mobile applications use 100s of libraries encompassing 1000s of functions each, with very deep cross-library calls, interspersed with jitted code. In addition, unlike conventional workloads represented by benchmarks such as SPEC CPU~\cite{spec} and GeekBench~\cite{geekbench}, a long tail is observed in modern mobile workloads on the {\it PC repeat distance}; i.e., the number of unique instructions executed between two occurrences of the same PC.
This causes many capacity misses in the CPU frontend structures such as instruction cache, TLB, and branch predictors, and results in starvation of the backend such that typical IPC is often below a quarter of the ideal value. We focus on I-cache misses in this paper; the idea can also be extended to other structures.

Software instruction-prefetching, either compiler-based~\cite{asmdb, coopprfm,kaynak2015confluence} or co-designed \cite{ispy}, tries to address that problem, but incurs often wasteful instruction overheads that are prohibitively large especially for prefetching on the return-path from call-stacks. Furthermore, the big-little structures often employed in mobile SoCs make it even harder to tune such approaches for timeliness, because cache structures and sharing differ among the big/middle/little cores~\cite{Exynos5,Qualcomm808,armbiglittle}.
Yet another problem is cross-library calls that are observed every few hundred instructions; compilers can optimize {\it only within} a library.
Fetch-directed instruction prefetching and its variants~\cite{reinman1999fetch,ishii2021re} also frequently fails on these workloads since they rely on branch prediction accuracy which itself drops severely due to large branch footprint and long repeat distances. 
Record-and-replay prefetchers \cite{ainsworth2024triangel,armcmc1, armcmc2, wu2019temporal} show partial success, but at the cost of hundreds of kilobytes of on-chip storage and the associated energy consumption.

We propose DEER, a deep runahead mechanism for prefetching instruction cache lines based on a tight collaboration between hardware and software. Our approach relies on software (compiler or binary analyzer) to generate metadata that encode the stable execution paths both within and cross libraries from the workload profile data. We implement multiple optimizations to ensure that the metadata is concise, accurate, and expressive of the most important aspects of the execution flow (e.g., control flow, loops, recursions, calls and returns). A hardware engine then employs this metadata as well as the run-time control flow information such as a Return Address Stack (RAS) to look far ahead of the normal program execution, and prefetches the instruction cache lines in a timely fashion. DEER is capable of continuously correcting itself by swiftly reacting to the mispredictions in the runahead and getting the metadata of the new path and resetting the runahead pathsaccordingly. Compared to conventional hardware-only prefetching approaches that are based on the record-and-replay method, DEER offers significantly higher agility, prediction accuracy and prefetching timeliness. Also, by leveraging the software-procured metadata, DEER dramatically reduces the size of on-chip storage and the energy which is required for training the hardware-only predictors. 

Despite ubiquity of mobile phones, little is available in the public domain on their workloads beyond benchmarks such as Geekbench~\cite{geekbench}, which are still not very well representing modern mobile workloads. 
Consequently, to obtain representative workloads, we employed a proprietary in-house simpoint-capture mechanism (Section~\ref{experimets-setup}) and report experimental results on these simpoints; simpointing flow is well established and results are reproducible. 
gem5 simulation results confirm multi-fold gains over full-hardware rivals at a fraction of their on-chip storage requirement.

Our key contributions can be summarized as:

\begin{itemize}
 \item We provide a software-hardware co-designed instruction-prefetcher that delegates the training and analysis part of prefetching to an in-software profiling mechanism, effectively eliminating the hardware's need to large storage. Moreover, our profile analysis step breaks through loops and recursions to go deeper into the future instruction stream.

\item We propose a simple SW/HW interface with small context-switch overhead; only a single system register is saved/restored.  Moreover, our choices of prediction-trigger points as well as the metadata granule (the HyperBlock) hit a good balance between accuracy, metadata overhead, and the frequency of prediction invocations. We provide details of the micro-architecture model as well as the metadata preparation process.

\item Our evaluation is on real-world modern workloads captured from real applications execution on the phone. This is important since publicly available benchmarks are far from representing the demanding current mobile applications.

 \end{itemize}
\section{Motivation}
\label{motivation}

First, we demonstrate a characteristic feature, \emph{long PC-repeat distance}; 
and show that it is pronounced more severely in mobile workloads, differentiating them from conventional benchmarks. It is also a root cause of poor reuse in I-cache (and other hardware-tables) that makes hardware-scaling an unattractive alternative.
We then quantify how much gain could be obtained if an oracle runahead prefetcher could predict all long-latency I-misses and prefetch all of the upcoming instructions into L2 cache. 
Finally, we show how a simple scheme allows to predict future cachelines in paths with high accuracy. Since the application behavior is fairly stable and follows a higher level pattern; this is the scheme that we employ in this paper.

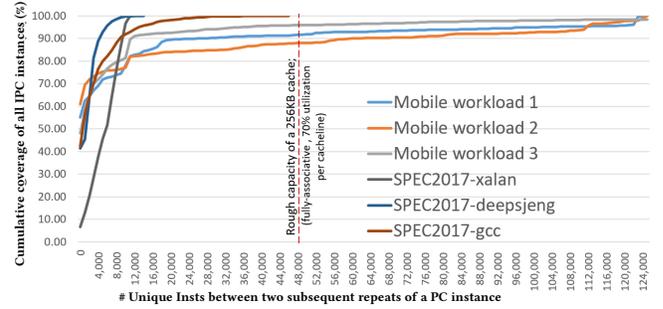
\begin{figure}
    \centering
    \includegraphics[clip=true, trim=1.1cm 1cm 0 0, width=1\linewidth]{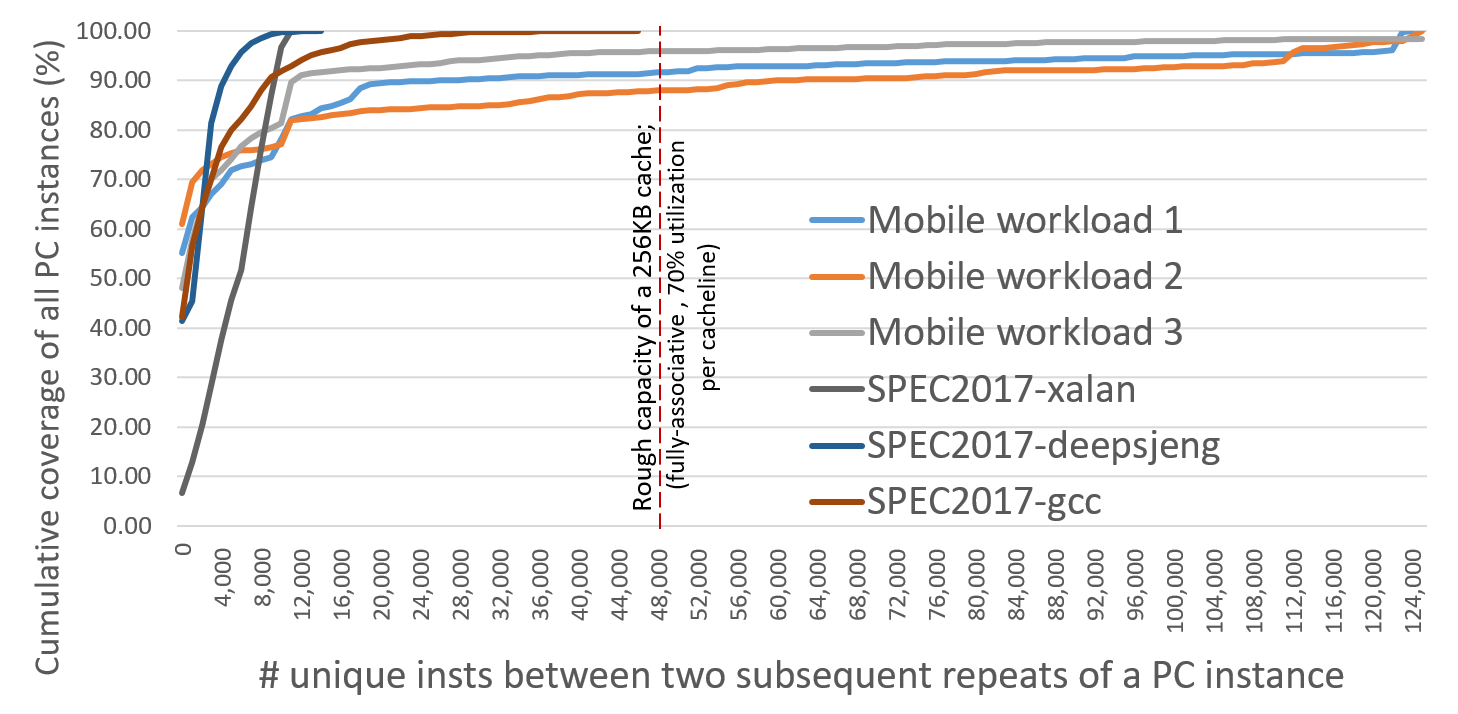}    
    \caption{ \small Repeat distance of PC instances; Mobile applications exhibit a far longer tail than SPEC workloads. 
    }
      \setlength{\unitlength}{1cm}
    \begin{picture}(0,0)
        \put(-4.3,1.9){\makebox(0,0)[b]{\rotatebox{90}{\textbf{\tiny \parbox{7cm} {Cumulative coverage of all IPC instances (\%)}}}}}
        \put(1,1.4){\makebox(0,0)[b]{\textbf{\tiny \parbox{8cm} {\# Unique Insts between two subsequent repeats of a PC instance}}}}
    \end{picture}
    
    \label{fig:reuse-distance}
  \end{figure}

To evaluate the \emph{repeat distance}, we measured the number of unique instructions executed between two subsequent instances of each PC in a number of mobile workloads and compared them against that of SPEC2017; Figure~\ref{fig:reuse-distance} shows a select number of them but others also follow a similar pattern. 
The figure shows that mobile workloads have 3-6X longer tails going beyond the capacity of industry's largest L1-I caches~\cite{intel-lunar-lake,qualcomm-oryon}: 5 to 15\% of PCs have reuse distances larger than 48K instructions; this would fully sweep a 256KB fully-associative cache of 70\% utilized cachelines.
Consequently, capacity misses occur more often, and reuse of the instructions in L1-I is less frequent. This calls for better prefetching mechanisms to reduce the corresponding  misses. 

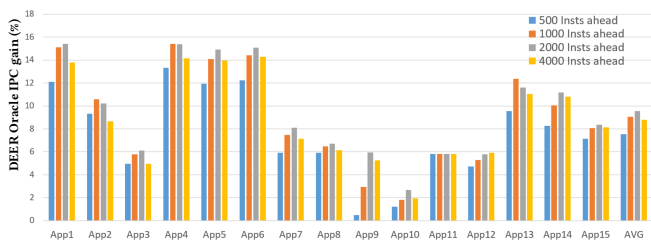
\begin{figure}
    \centering
    \includegraphics[width=1\linewidth]{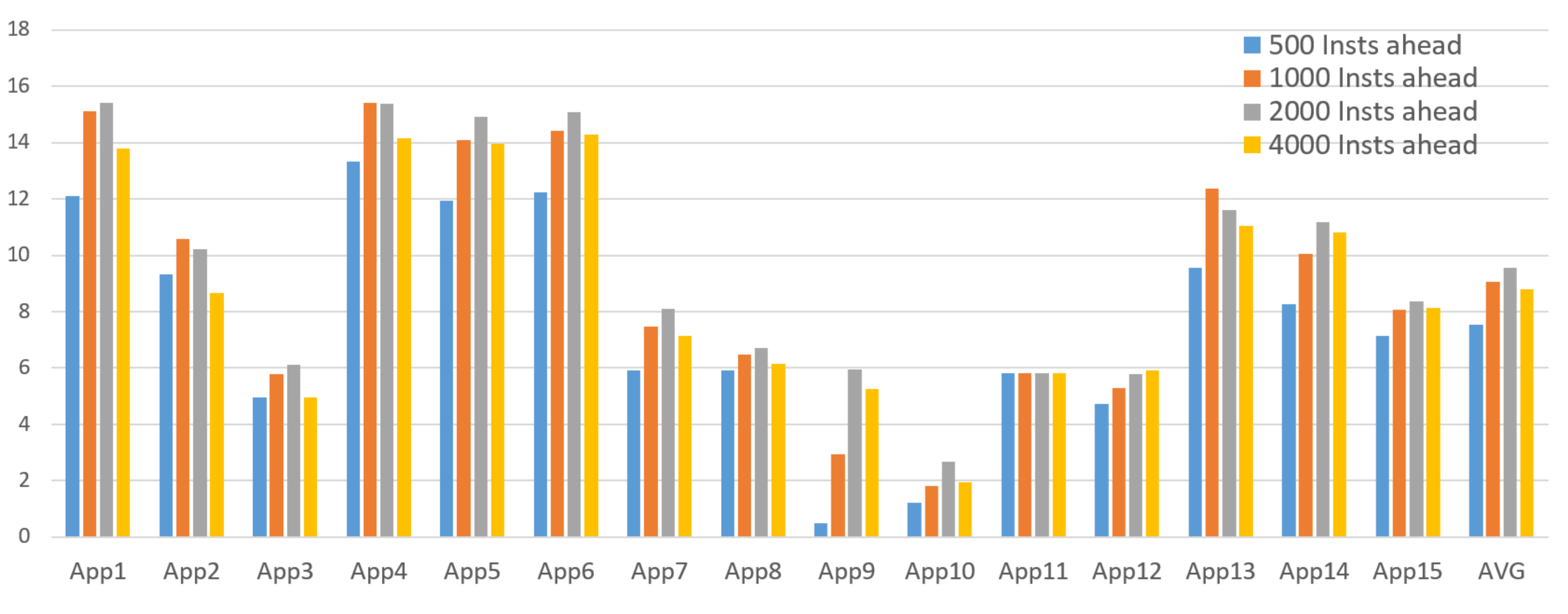}
    \caption{\small IPC gains by an oracle N-instructions-ahead prefetcher for different prefetch distances.}
     \setlength{\unitlength}{1cm}
    \begin{picture}(0,0)
        \put(-4.3,2.5){\makebox(0,0)[b]{\rotatebox{90}{\textbf{\tiny \parbox{5cm} {DEER Oracle IPC gain (\%)}}}}}
    \end{picture}
    \label{fig:oracle}
 \end{figure}

To evaluate the estimated upper bound of possible gains by such a prefetching scheme, we implemented an oracle runahead scheme that prefetches the exact instruction cachelines that will be executed {\it N} instructions later; prefetch is to the L2 cache and system settings are same as Section~\ref{experimets-setup}. Figure~\ref{fig:oracle} confirms that a large opportunity exists here.
Gains come from reduced I-cache misses provided by the 100\% accurate oracle; prefetch into L2 reduces sensitivity to timeliness as well.

Finally, we also observe that a reasonably stable pattern exists in the trace of instruction cachelines of our mobile workloads, and an extended most-likely-successor (MLS) scheme---Section~\ref{design}---can fairly well predict that pattern.  
As a measure of stability of this coarser grained behavior of applications, we measured the intersection-over-union (IOU) for the two sets below, computed per PC instance: 

\begin{itemize}
    \item $S_{PC}^{Pr}$: The set of unique cachelines {\it predicted} by the MLS scheme to be executed after that PC. The set formation stops upon covering N dynamic instructions.
    \item $S_{PC}^{Ex}$: The set of unique cachelines, with same size as $S_{PC}^{Pr}$, that we see {\it executed} after the same trigger PC.
\end{itemize}

\begin{figure}
    \centering
    \includegraphics[width=1\linewidth]{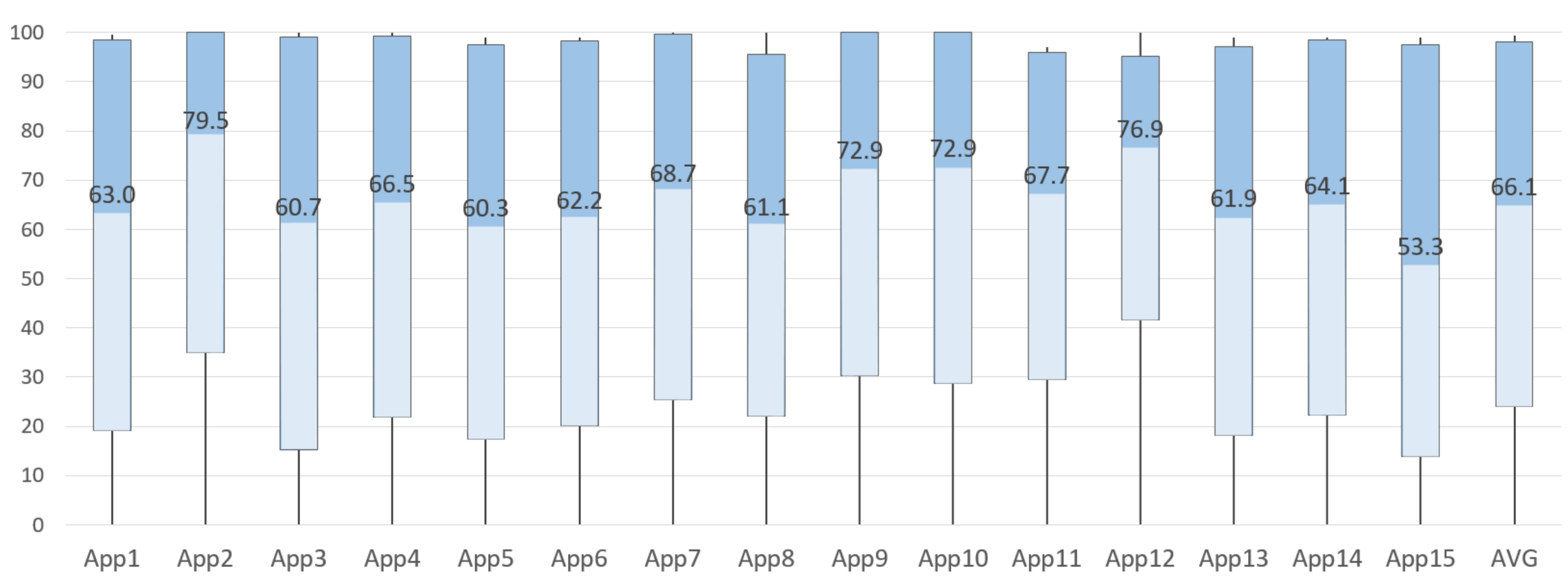}
    \caption{\small Box-and-whisker plot for path prediction accuracy of the MLS (most likely successor) scheme.}
     \setlength{\unitlength}{1cm}
    \begin{picture}(0,0)
        \put(-4.3,2){\makebox(0,0)[b]{\rotatebox{90}{\textbf{\tiny \parbox{5cm} {DEER Prediction accuracy (\%)}}}}}
        
    \end{picture}
    \label{fig:accuracy}
 \end{figure}
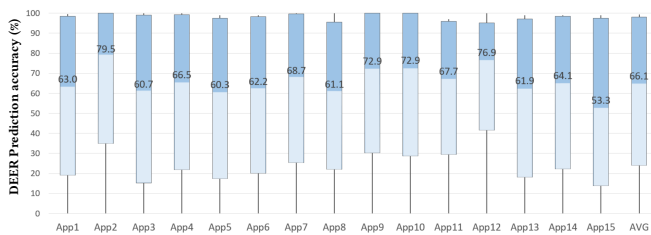

Figure~\ref{fig:accuracy} shows the span of the values for IOU of $S_{PC}^{Pr}$ and $S_{PC}^{Ex}$ sets, over all PC instances for each application; on average, the extended MLS scheme provides 67\% correct prediction of the unique cachelines that will be executed later.  
We show that this accuracy is enough to cover many of the long-latency I-misses in an inexpensive co-designed I-prefetcher. 
We also studied other enhancements, such as call-context-awareness, to improve the accuracy of the predicted paths; however, since they require extra metadata storage, in this paper we focus on a simple most-likely-successor scheme to keep a balance between accuracy and overhead.

\section{Related work}
\subsection{Software-Only Mechanisms}
Software prefetching, regardless of inserted manually by the programmer or automatically through a compiler, faces difficulties in timeliness, accuracy and coverage as program dynamic behaviour is unknown at coding/compile time~\cite{chacon2023characterization}. 
APT-GET ~\cite{jamilan2022apt} aims to ensure the timeliness of software prefetches by determining the prefetch distance and injection site.
Prior work~\cite{asmdb, coopprfm} employ PGO to improve accuracy, 
however the additional dynamic instruction overhead, inherent in software prefetching, takes away from the gains of fewer instruction cache misses, assuming even perfect timeliness and accuracy. 
A further difficulty in mobile workloads comes from the big-little structure employed in virutally all mobile SoCs; each big/middle/little core typically has different levels and sizes of cache, and a thread at runtime may be rescheduled to any of those cores without the application ever noticing. This makes it harder to statically tune the software prefetching.
Other shortcomings of compiler-only approaches are the inability to put cross-library prefetches, and the high instruction-cost of prefetching on the return-path.
DEER does not insert prefetch instructions; but adjusts a hardware pointer to the location of the meta-data in main memory. 
Big-little architecture also is not a problem for DEER since it is the run-time host core that initiates the prefetch.
Further, DEER covers both cross-library and return-path prefetching as the formed metadata crosses library boundaries and a return-address stack in the hardware is employed respectively.

\subsection{Hardware-Only Mechanisms}
Fetch-directed instruction prefetching (FDIP)~\cite{reinman1999fetch} is a common technique employed in most modern processors~\cite{rupley2018samsung,pellegrini2020arm}; it employs the branch prediction complex to predict branches further ahead of current fetch, and prefetches them. FDIP also performs poorly when branch predictors degrade due to large branch footprint and long repeat distance in the workload.
Similarly~\cite{srinivasan2001branch} employs branch predictors to predict instruction cache misses, while others~\cite{pierce1996wrong} propose prefetching along the wrong path; all such techniques face same above issue when branch prediction accuracy drops.

Record and Replay (RnR) techniques~\cite{ignite, cache_restore, recap, ferdman2011proactive, pellegrini2021arm, ferdman2008temporal} address some drawbacks of software prefetching by learning dynamic access patterns of instructions, and replaying them at a later time. 
They provide high coverage of misses by spanning over large execution windows, but impose large storage overheads for the recordings. Several advancements on RnR instruction prefetchers reduce storage overheads by eliminating redundancy in the recordings~\cite{ansari2022mana} or improve timeliness and usefulness of the prefetches by adaptively adjusting depth of prefetch~\cite{oh2024udp}, however they still require complex logic for recording and replaying.

UDP ~\cite{oh2024udp} adaptively adjusts FDIP depth and wrong-path prefetches to improve depth and usefulness of the prefetches to improve cache pollution and accuracy. 
~\cite{ansari2020divide} addresses sequential misses, discontinuity misses, and BTB misses, and attempts to optimize performance by prefetching across discontinuity caused by branches.  
Boomerang~\cite{kumar2017boomerang} attempts to enhance the BTB performance by introducing small additional hardware.
PDIP ~\cite{godala2024pdip} improves FDIP by identifying the mispredicted branches and using them as trigger points to prefetch for misses that FDIP had failed to cover.
MANA ~\cite{ansari2022mana} identifies spatial correlations among instruction cachelines and records and replays those regions. 
RDIP ~\cite{kolli2013rdip} uses return-address-stack (RAS) as the program context which can be used to track patterns and predict future instruction misses by inspecting the execution history.
The authors in \cite{nesbit2004data} take advantage of the global history buffer and the spatio-temporal locality of the executed code. Shift~\cite{kaynak2013shift} shares the instruction history across cores to enhance the performance of multi threaded programs more efficiently by embedding the history in the LLC.

A fundamental problem with these full-hardware mechanisms is the storage requirement; since the repeat-distance of many PCs are large (see Section~\ref{motivation}), the recorded information has to be stored for a long time before replay. 

The MLS mechanism in DEER can be viewed as an offline compute---as opposed to online store and deduce---of meta-data, and thus it significantly reduces the amount of on-chip storage needed.

\subsection{Co-Designed Mechanisms}
\label{section:codesigns}
Co-designed mechanisms employ changes on both the software and the hardware sides to prefetch instructions~\cite{ispy}~\cite{efetch} or preload BTB entries~\cite{khan2021twig}.
Since DEER is also a co-designed  technique, we use I-SPY and Efetch as co-designed competitors described below.

\textbf{I-SPY}~\cite{ispy} uses x86 PEBS \cite{doweck2017inside} to collect I-miss profile and then employs LBR \cite{lbr} to create the dynamic CFG of BBs leading to the missed BB. Targeting BBs that most likely lead to the missed BB, and injecting I-prefetch instructions there. Coalescing of multiple I-prefetches as well as prefetching conditionally based on the call-context are also added. I-SPY must stay conservative due to the instruction overhead.
In contrast, we do not inject prefetch instructions into the code and leave this to the hardware (to I-prefetch based on the metadata). Further, it is expensive for I-SPY to inject prefetches for the return-path from callee functions to callers since a callee may have several callers; we employ a RAS for that. 
I-SPY also faces challenge to find proper injection sites for misses located after loops or recursions, but our cycle-skipping mechanism addresses them.

\textbf{EFetch}~\cite{efetch} observes that in event-driven interpreted code such as Java Script, often the external events along with the call-context, determine the callee function and the part of its body to be executed. They modify the browser source code to produce an event signature and pass it to the hardware which records and correlates that event as well as the call-context to the executed code of the callees for future replay. They decide that 3 levels of call-context and prefetch of 1 level of callees is enough for the benchmarks. This parameter setting is not a good choice for mobile apps where call stacks get to 100s of calls deep and repeat distances are very long; DEER takes into account the deep, large, and sparingly executed call stacks since it gradually prefetches the most-likely next set of callees as well as return paths.

\textbf{Hierarchical Prefetching}~\cite{zhang2025hierarchical} Similar to the work presented in this paper, ~\cite{zhang2025hierarchical} uses coarse grained sections of code, called \textit{Bundles}, and passes only the Bundle starting points to the hardware by introducing new instructions (or using the reserved bits if the ISA allows it). However, these Bundle headers are used only to record coarse traces of program execution and cachelines and replay them upon the next sighting of the same Bundle start pc. The boundaries of a bundle are defined by a size threshold. 
\section{Design}
\label{design}

\begin{figure*}
    \centering
    \includegraphics[  width=1\linewidth]{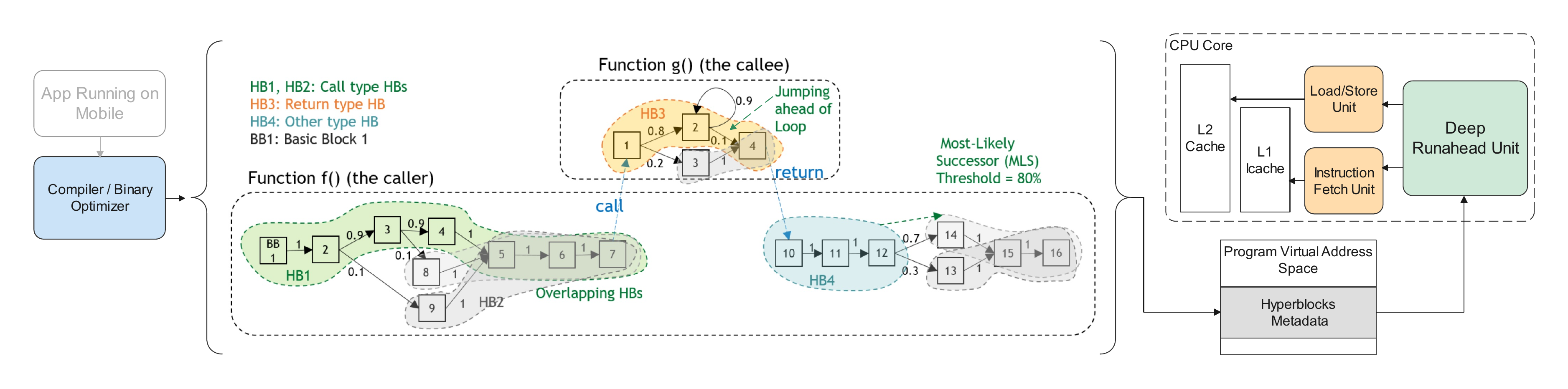}
    \caption{\small DEER system overview: deep runahead with software-generated metadata.}
    \label{fig:overview}
\end{figure*}

We use the terms in Table ~\ref{tab:terms} to describe the system.

\subsection{System Overview}
\label{section:system-overview}

\begin{table}[t!]
\footnotesize
\begin{center}
\caption{\small The terminology used in the paper.} 
\label{tab:terms}
\begin{tabular}{|p{1.5cm}|p{6.2cm}|}
\hline

\textbf{HB} & {\it Hyperblock}, is a chain of BBs \emph{within a function} executed with a proability above a threshold; e.g. in Figure ~\ref{fig:overview} HB4 is cut at BB12 as the branch probability (0.7) is below the threshold (0.8). Tails may overlap (e.g. HB1 \& HB2) but each start-PC is unique.\\
\hline
\textbf{HB PC / HB-Start PC} & An HB is identified by the PC of its first instruction; thus we use HB and HB-PC interchangeably. \\
\hline
\textbf{Trigger PC}& The target-PC of a call/ret; that would designate an HB (every Trigger PC is an HB PC), and sets off an MLS guided prefetch.\\

\hline
\textbf{HB type} & The tail-end branch of an HB defines its type; it can be {\it call}, or {\it return}, or {\it other}. \\
\hline
\textbf{MLS} & {\it Most Likely Successor}, is defined per HB (and BB), and  designates the HB (BB) that is most often executed immediately after it according to the examined profile.\\
\hline
\textbf{next-HB} & The MLS HB for a given HB. In Figure ~\ref{fig:overview} the next-HB of HB1 is HB3 even though HB1 might have other possible successors.\\
\hline
\textbf{Runahead depth}& The distance the MLS chain aims to achieve, in number of HBs, or instructions, contained in the chain. \\

\hline
\textbf{HB-Metadata} & Complementary information per HB, such as its enclosing cache lines, its MLS HB, and its type (See Figure ~\ref{fig:ssra})\\
\hline
\textbf{Metadata table} & The array of all metadata entries (also called {\it metadata lines}) in memory.\\ 
\hline
\textbf{Metadata-cache} & If employed (e.g. in dynamic runahead, Section~\ref{sec:dyn-runahead}), stores the most recently used metadata entries for reuse. \\
\hline
\textbf{RAS} & A {\it Return-Address Stack} to keep track of the call stack.\\

\hline
\end{tabular}
\end{center}
\end{table}

Figure~\ref{fig:overview} demonstrates an overview of the DEER system. At the high level, a compiler or binary optimizer extracts the HB information from the program control flow and call graph that is augmented with profile data. The profile data contains information about the frequency of execution for each basic blocks as well as the probabilities for branches and call targets (for the indirect calls). Each HB represents a stable unit of execution within a function in the control flow graph; i.e., a collection of basic blocks that are expected to run in sequence based on the profile data. The profile data contains key information about the frequency of execution for basic blocks and the probabilities of the branch executions both in terms of their directions as well as targets.  The HB information are encoded into a {\it Metadata Table} that is loaded as metadata section in the program address space. The starting address of this metadata region is passed to the CPU core which is augmented with a {\it Deep Runahead Unit (DRU)} which is responsible to track program execution and search for the HB accordingly. The HB information is streamed on demand into the DRU continually as the program execution progresses forward: 
At certain points in the execution, such as the commit of a call/return instruction, a set of the upcoming HBs are identified by the DRU and their enclosing cacheline addresses are passed to the Instruction-Fetch Unit or Load/Store Unit to be prefetched either into the L1 Instruction Cache or the unified L2 cache respectively.

\subsection{Deep Runahead Methods}
A simple {MLS} scheme is the building-block we employ to predict future execution paths.
The HBs are formed separately per function, and another MLS scheme is applied on top of them to form cross-function most-likely paths.
Upon commit of a call/return instruction, its target PC is used as the {\it trigger PC} to start forming a runahead-chain of HBs. 
We discuss two runahead mechanisms: (i) {\it dynamic runahead} can go deeper into the future, but needs multiple referrals to the metadata table, and (ii) {\it semi-static runahead} employs only one or two accesses to (a modified) metadata table per invocation, but would go a shallower depth instead.

\subsubsection{Dynamic Runahead}
\label{sec:dyn-runahead}

Algorithm~\ref{algo:mls} shows the dynamic runahead method based on the MLS information. During normal execution of the program, the
{\it trigger PCs} are identified, and then
the dynamic runahead mechanism attempts to predict the upcoming instructions and prefetch their corresponding cache lines into the caches. 
In this paper, we often aim for a depth of 50 HBs ahead, although this number is configurable. An HB can be of type $call$, $return$, or $other$ depending on its tail-end branch. 
The prediction mechanism simply forms a dynamic chain of $<HB,Next HB>$ links, where {\it Next HB} is the MLS of the current HB if HB is of type $call$ or $other$ (lines 7 and 9); otherwise, {\it Next HB} is read from RAS since HB type is $return$ (line 12).  For example in Figure ~\ref{fig:overview}, upon seeing HB1, the address of HB4 is pushed on RAS, and next-HB is predicted to be HB3. 
For the next step of the chain formation, the next-HB of HB3 (which is a $return$ HB) is read from RAS (thus HB4). This dynamic process shown in Algorithm~\ref{algo:mls} stops if either the depth of the chain reaches the set cap (line 2), or the next-HB is not found in the metadata-cache (lines 14-17).

The upper right part of Figure~\ref{fig:ssra} shows an example of metadata for dynamic runahead; 
per HB, the metadata line contains the most likely next-HB, HB type, return address (for $call$ HBs only), and the list of cachelines enclosing the HB.
For a dynamic runahead depth of 5 HBs ahead, when the call/return instruction leading to the start-PC of HB1 is committed, the chain of HB2, HB3, and HB4 are predicted as next-HBs, while HB6 and HB5 are respectively pushed on the RAS. The next-HB of HB4 is then read from RAS as HB5, and thus the cachelines enclosing HB5, designated as x5 in the table, will be prefetched. 

Experiments showed us that this simple MLS-based prediction mechanism can yield good accuracy in identifying the upcoming HBs. However, this approach requires the in-hardware HB predictor to have a rather large metadata-cache to achieve good performance simply because the $<HB, Next HB>$ lookups could exhibit a random memory access pattern in the metadata; moreover, $N$ (say 50) lookup operations into the metadata cache to traverse a chain of $N$ HBs takes multiple cycles and imposes complexities in handling multiple chains simultaneously.

\begin{algorithm} [t]
\small
\caption{Dynamic Runahead Logic}
\raggedright \textbf{System settings:} Runahead-depth  \\
\raggedright \textbf{Input:} Trigger-PC, the start PC to form the chain of HBs. \\
\textbf{Output:} List of I-cachelines to prefetch, and metadata lines to fetch from memory system to metadata-cache.
\begin{algorithmic}[1]
    \STATE curr-HB = Trigger-PC; $entry$ = EMPTY
    \FOR{$depth \gets 1$ to Runahead-depth}
        \IF{ curr-HB in \textit{metadata-cache}} 
            \STATE $entry$ = metadata-cache[ curr-HB ]
            \IF { $entry$.HBtype is $call$ }
                \STATE push $entry$.return-address to RAS
                \STATE curr-HB = $entry$.nextHB
            \ELSIF { $entry$.HBtype is $other$ }
                \STATE curr-HB = $entry$.nextHB
            \ELSE
                \STATE  {\color{gray} // $entry$.HBtype is $return$}
                \STATE curr-HB = pop from RAS
            \ENDIF
        \ELSE             
            \STATE Issue metadata-cache-miss for curr-HB \color{gray} // \textit{ MLS chain stops at any point where the HB is missed in the metadata-cache} \color{black}
            \STATE break
        \ENDIF
    \ENDFOR
    \RETURN $entry$.cachelines-list 
\end{algorithmic}
\label{algo:mls}
\end{algorithm}

\subsubsection{Semi-Static Runahead (SSRA)}
\label{sec:ssra}
\begin{figure}[t!]
    \centering
    \includegraphics[width=1\linewidth]{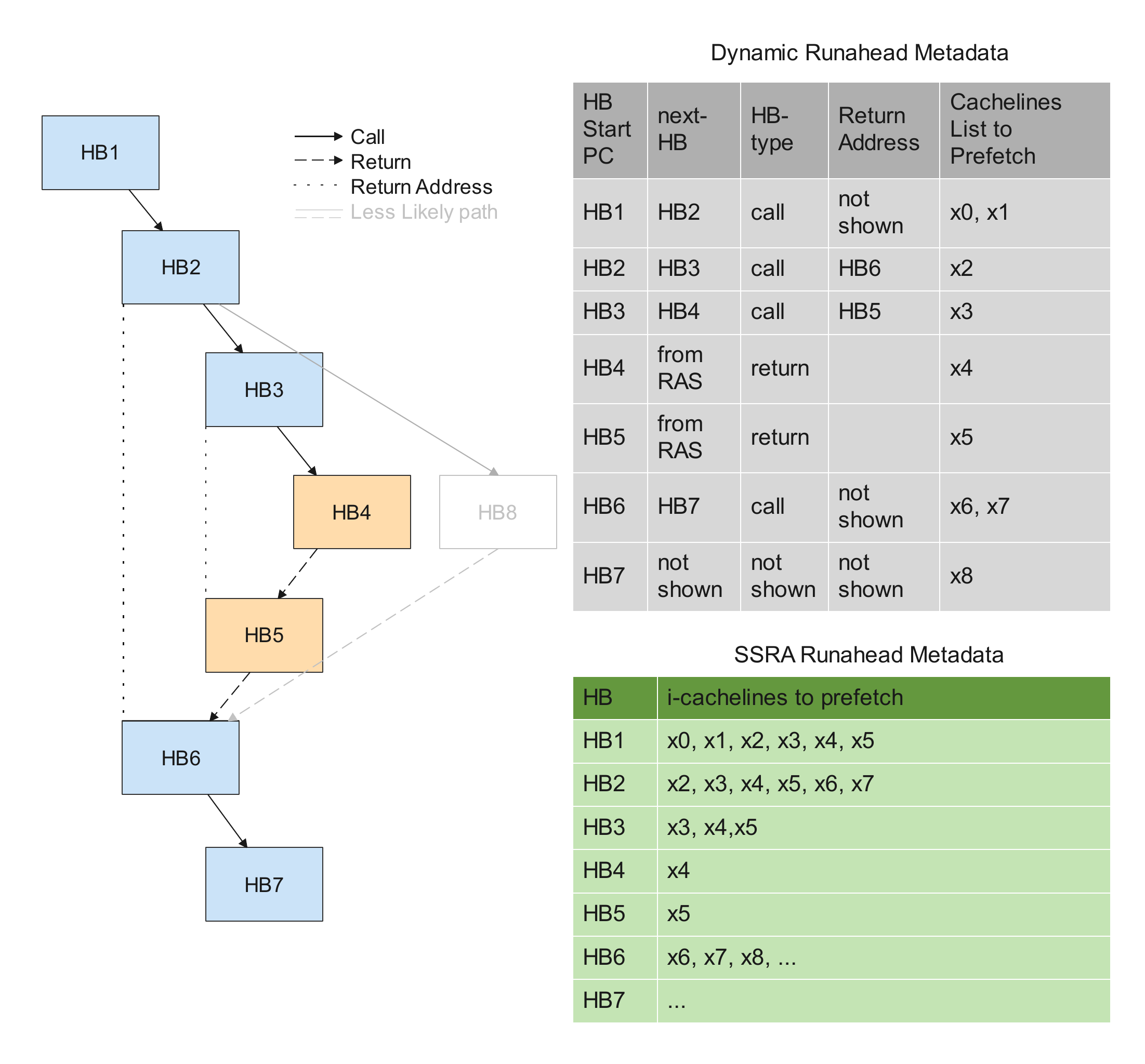}
    \caption{\small An example of HB metadata for dynamic as well as SSRA runahead modes for a runahead depth of \textbf{5} HBs. 
    }
    \label{fig:ssra}
 \end{figure}

To address the above issues of the dynamic runahead approach, we use the {\it Semi-Static Runahead}, or $SSRA$, that statically (before run-time) computes the runahead chain as far as possible. In this setup, the metadata of each HB gives the entire list of I-cachelines in its MLS chain; this list of cache lines is then prefetched 
upon invocation of the runahead mechanism.
The algorithem to calculate a SSRA chain is quite similar to Algorithm~\ref{algo:mls} with two main differences: 
1) there is no {\it metadata-cache miss} since metadata conversion to SSRA variant is done statically,
2) instead, a new limitation is the {\it statically-formed RAS}, which consequently does not contain the caller information beyond the starting function.
For example in Figure ~\ref{fig:ssra}, when trying to form the SSRA chain for HB3 statically, it is not known whether HB2 was the caller and HB6 is the return-address, or the caller was indeed someone else; consequently, the algorithm has to stop the static runahead at HB5. 
Accordingly, in the SSRA runahead metadata table in Figure~\ref{fig:ssra}, the HB3 line tells to only prefetch the cachelines of HB3, HB4, and HB5.

Handling of $other$-type HBs is also similar to Algorithm~\ref{algo:mls}; the SSRA chain simply includes the next-HB in the chain.
Note how this static chain formation removes the need for metadata-cache since the multi-step mapping is removed. 
Consequently, the hardware also becomes very simple---see Algorithm~\ref{algo:ssra}.
For the rest of the paper, our proposed hardware and experimental results focus on the SSRA method unless stated otherwise.

\subsection{Hardware Overview}
\label{section:hardware}
    
\begin{figure}
    \centering
    \includegraphics[width=1\linewidth]{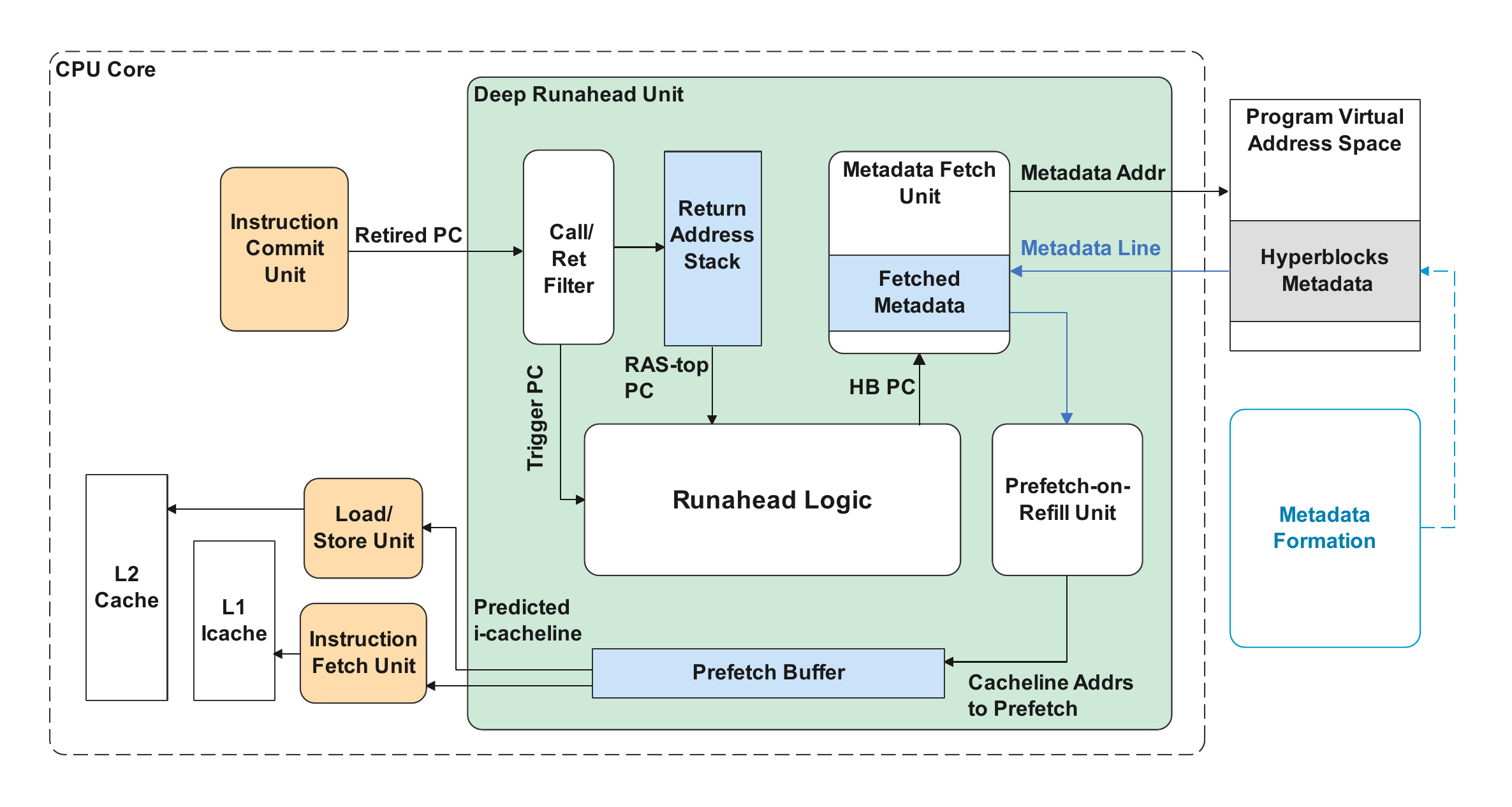}
    \caption{\small Hardware block diagram of DEER prefetcher.}
    \label{fig:hw}
 \end{figure}

The invocation of DEER prefetching is non-speculative; i.e. the mechanism is engaged and the prefetch-list is decided based on the latest {\it committed} instruction, as opposed to a {\it fetched} instruction. 
Figure \ref{fig:hw} shows the hardware unit of DEER prefetcher. 
First, the committed instructions are filtered in the {\it Call/Ret filter} so that only call/return instructions are considered. 
The {\it Return Address Stack} is also updated here by pushing the return-address upon committing a $call$ instruction, and popping it for $return$ instructions. The PC of the target of that call/return is also obtained as the start of the runahead chain; we call it the {\it trigger-PC}---see Table ~\ref{tab:terms}.
Next, the {\it Runahead Logic} generates two memory-requests for the SSRA metadata lines that contain the list of cache lines to prefetch: one request for the {\it trigger-PC} and another one for {\it top-of-RAS PC}---see Algorithm~\ref{algo:ssra}. 
These requests are handled by the {\it Metadata Fetch unit}.
These metadata-request packets use regular memory-access paths and facilities, and do not need additional dedicated resources.

There is no metadata-cache in Figure~\ref{fig:hw} as justified in Section~\ref{sec:ssra}.
When the metadata-requests are back from memory system, {\it prefetch-on-refill Unit} 
extracts the list of cache lines in each, and pushes them to the tail of {\it Prefetch Buffer}; this is the \textbf{prefetch-on-refill} feature in the lower part of Algorithm~\ref{algo:ssra}. 
The second feature, \textbf{RAS-top prefetch} (line 2 of Algorithm~\ref{algo:ssra}), is  a complementary operation that issues another metadata-request, this time for the PC on top of the RAS, in an attempt to regain some of the runahead depth lost in the SSRA formation.
Note how there is no guarantee that the HB corresponding to the RAS-top-PC is a direct continuation of the trigger-PC's HB; such case would create a gap between the two prefetched chains. Consequently in general, the RAS-top-prefetch could be too early in such a case and may pollute the cache accordingly, but the experimental results showed that in practice, it improves the gains by 9.6\% on average across all apps.

\begin{algorithm}
\small
\caption{SSRA Runahead Logic}
\raggedright \textbf{Input:} Trigger-PC, the start PC to get the chain of HBs. \\
\textbf{Output:} Metadata requests to fetch from memory system.
\begin{algorithmic}[1]
    \STATE Issue metadata-request for Trigger-PC
    \STATE Issue metadata-request for top-of-RAS PC {\color{gray} \it // \textbf{~RAS-top- prefetch} feature}
\rule{\linewidth}{0.1mm} 
\textbf{\\Prefetch-on-refill feature:} {\color{gray} \it // In parallel to the above}
    \STATE $entry$ = metadata line arrived from memory system
    \STATE Push $entry$.cachelines-list to prefetch-buffer tail
\end{algorithmic}
\label{algo:ssra}
\end{algorithm}

Note how the SSRA-formation mechanism helped to simplify the hardware compared to the dynamic runahead in Algorithm~\ref{algo:mls}; 
This is achieved because most of the chain-chasing is now done statically. 
The downside of the SSRA scheme is that the HB-chain is cut shorter than the dynamic counterpart, resulting in some gain loss.
Experiments show each SSRA chain covers 15.5 HBs on average across all apps (the effective runahead depth in \#instructions is shown in Table~\ref{tab:depth}).
Nevertheless, the obtained storage saving and hardware simplicity proved more valuable than the lost gain.

\subsection{Software/Hardware Interface}
\label{sec:sw-hw-interface}
Figure~\ref{fig:hb_tab} describes the encoding details of the entries in the metadata table in memory. Each entry in the metadata table is broken into two 64-bit subentries (group1 and group2, each spanning one row in the figure),
each of which holds the address and cacheline-selector bitmap for three 512-byte {\it regions} (8 cache lines per region).
For each region, we encode a 30-bit {\it base address} followed by an 8-bit bitmap that denotes which cache lines in the region are included in the SSRA chain; remaining upper bits come from the HB PC.

For the first region in each 64-bit group, the base address is encoded in full 30\-bit form. For the next two regions in the group, the base addresses are calculated using 5-bit deltas with 512-byte granularity. As a result, the three regions for each group can be up to 16-KBytes apart from each other, each encoding the addresses for 8 cache lines (24 in total). 
Thus in total, each HB metadata entry consumes 16 Bytes and can encode addresses for up to 48 instruction cache lines. 

To load the metadata line corresponding to a given HB, 
the offset of the entry in the metadata table is obtained by hardware by computing a hash value of the given HB PC.
The per-process base address of the HB metadata table is loaded in a system register, {\it HBT\_PTR}, upon either the program load time or during the restoring of registers at the end of each context switch. 
Start PC of the HB is sent by
{\it Runahead Logic} to {\it Metadata Fetch Unit} to compute the hash and get the metadata line from memory (Algorithm~\ref{algo:ssra} and Figure~\ref{fig:hw}).
Table~\ref{tab:benchmarks} shows the in-memory storage overhead for the metadata is around 9\% of the size of exercised code path, and only 2\% with respect to code plus data footprint. In our experience with analyzing many real world mobile applications and libraries, only less than 10\% of the program binary is actually exercised most of the time. As a result, the expected storage overhead for the HB metadata is negligible compared to the size of the program binaries. 
{Even compared to the \textit{memory-resident} code and data footprint of the apps, the metadata overhead is negligible; metadata adds only 2.17\% to the memory requirements of the app, which itself is not currently a top concern for phone manufacturers.

\begin{figure}
    \centering
    \includegraphics[width=0.95\linewidth]{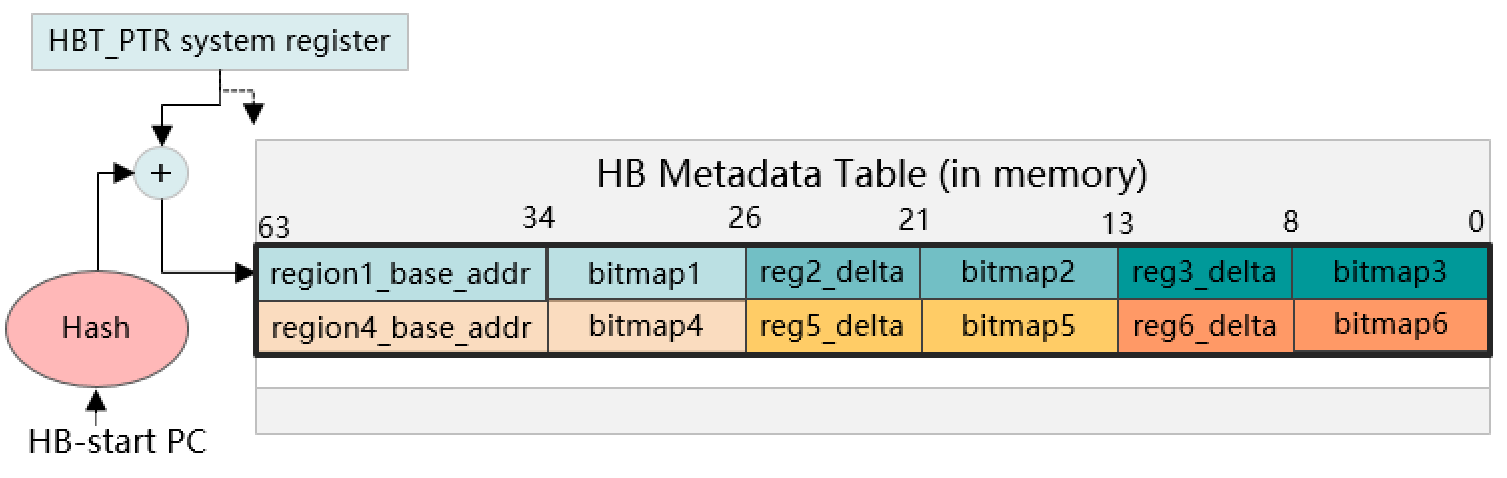}
    \caption{\small HB metadata encoding and access scheme using the metadata-table-pointer system register, $HBT\_PTR$.}
    \label{fig:hb_tab}
 \end{figure}

\subsection{Metadata Formation Method}

Figure~\ref{fig:metadata} shows the overall flow for generating the metadata for mobile applications. As the first step, a branch profiler obtains Taken-branch records from the CPU core's Performance Monitoring Unit (PMU) which is equipped with ARM's Branch Record Buffer Extension (BRBE)~\cite{brbe,arm2020,armarch}. 
A Path Profiler tool combines the branch probabilities with the information from the application's control flow graph (CFG) and call graph to produce HBs by creating highly probable branch paths using branch profiling information.These HBs are processed and linked together in several steps. 
First, for each HB, the most-likely successor HB is identified. Secondly, iterative execution patterns such as loops and recursions are identified in the program CFG to allow linking the HBs within the loops and recursions to the HBs that are executed after the exit from the loops or recursions---see the loop skipped over by HB3 in Figure~\ref{fig:overview} as an example; this is an important step that allows to {\it jump out} from such cycles, and go deeper into the forthcoming instructions during a runahead operation. 
Thirdly, the HB metadata is formed with the format explained in Section~\ref{sec:sw-hw-interface} by chaining both the HBs within functions and those across function calls. The metadata is then loaded into a non-cacheable area of the program address space (either at the application restart or in the live application image) and the HB-metadata Table Pointer {\it (HBT\_PTR)} system register is updated to point to the start address of the metadata table in memory.  
Similar to~\cite{stojkovicmosaic}, BRBE sampling is done online, but HB formation is run offline, outside of the mobile app critical path.

\begin{figure}[t]
    \centering
    \includegraphics[width=0.9\linewidth]{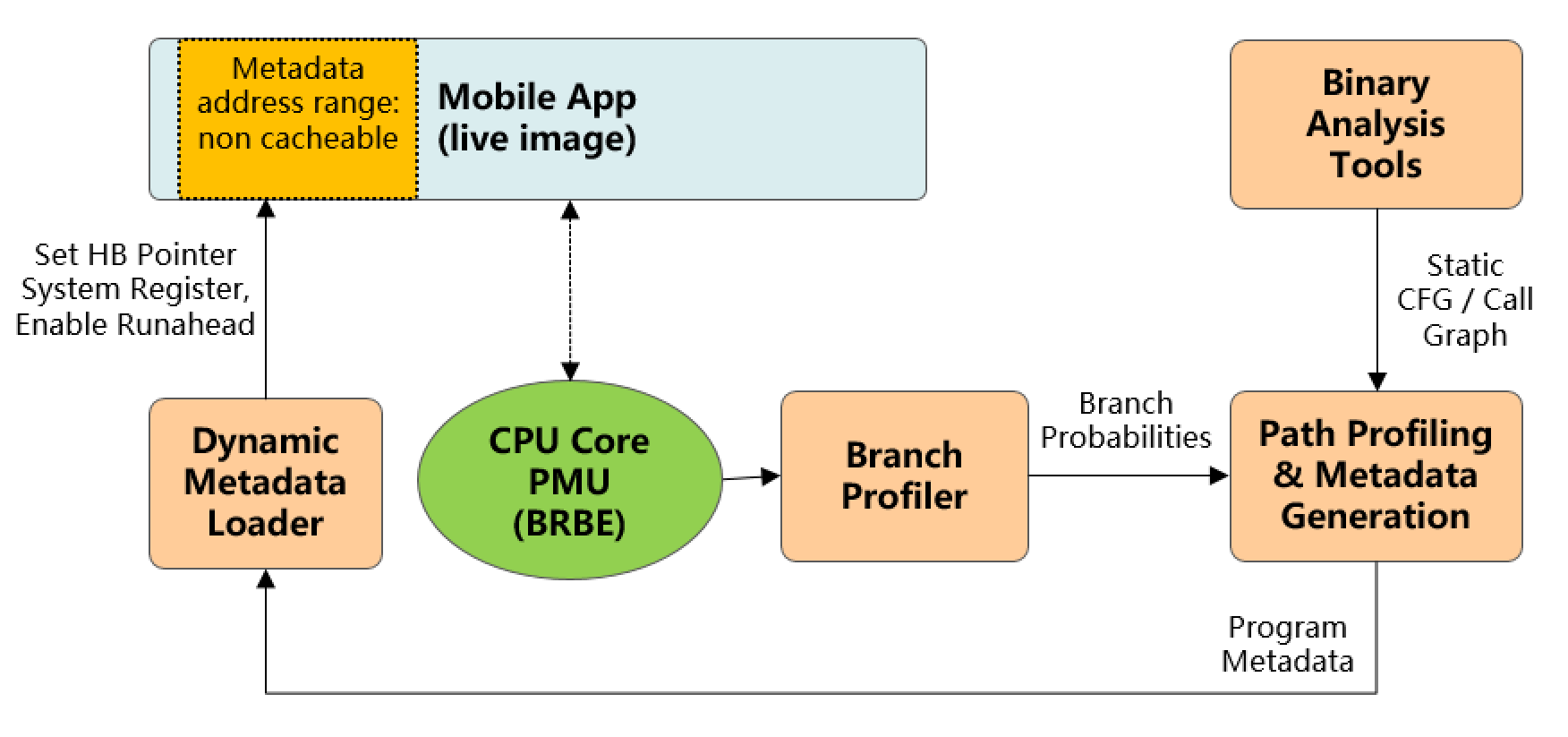}
    \caption{\small The metadata generation flow.}
    \label{fig:metadata}
  \end{figure}

\subsection{Discussion of choices and features}
\label{discussion}
\textbf{Choice of trigger PC:} 
Study of our mobile simpoints suite proved them function-call heavy; on average, there is a call instruction per every 50 instructions. This frequency is twice higher than a set of common server workloads we also examined. 
This led us to our choice of the trigger-PC: A call/return instruction is a frequent-enough trigger point.

\textbf{Choice of HB/metadata granule:} 
The above choice of trigger PC, gives also a natural choice for metadata granule in the SSRA method: the target PC of every call or return instruction designates one metadata entry. We further applied an optimization here by identifying HB-chains that are fully contained within only one other HB-chain; in that case the former metadata entry is removed since the only way to reach that HB-chain is through the latter, which obviously gets the former covered. These choices reduce metadata size to a marginal overhead as discussed in Section~\ref{sec:sw-hw-interface}.

\textbf{Self-correction mechanism:}
DEER continuously decides the prefetch list based on the \emph{retired} instruction; thus it corrects itself accordingly as the execution continues. For instance, imagine a change in Figure~\ref{fig:overview} where HB2 continues otherwise and ends in a different basic block that calls a different function $h()$ rather than the one shown in the figure. Now when the execution reaches HB1, the predicted path is as shown in the figure, but a bit later if the actual execution goes to HB2 instead, the prediction is corrected since the HB-chain now goes through the $h()$ function.
\section{Experimental Results}
\label{evaluation}

\subsection{Experiments Setup}
\label{experimets-setup}

To obtain real world workloads representative of modern mobile apps, we employed an in-house simpoint-capture mechanism; the target mobile app is run on the phone, and several instruction traces are captured from the designated processes and threads, using in-house  hardware equipment and software tools. Then similar to the clustering step in the simpoint methodology~\cite{1238020}, a number of the captured instruction-traces are chosen as representative simpoints for the app. Each such simpoint is an executable ELF with same exact instruction and data memory layout as the original mobile app, and executes same instructions as original. 
We evaluate DEER on 15 such above simpoints from various mobile apps.
The names of the apps are kept anonymous for business reasons; all of them are recent apps from among news and internet browsers (App1, App2), video games (App4, App5, App6, App15), video players (App13, App14), social networks (App3, App7), video call and payment apps (App11, App12), news readers (App9, App10), online shopping (App8) and the like . 

Each simpoint is the instruction sequence in a single thread of the mobile app, and encompasses system-libraries (70-80\% of the trace) and JITted code resulted from app-specific logic. User usage scenarios are studied and analyzed by business teams to identify typical user inputs and actions, which are then used when collecting the simpoints; this ensures representativeness of the simpoints for typical users and use cases. Moreover, each simpoint corresponds to specific usage scenarios; e.g. launching the app, opening the gallery, swiping pages, etc. 
Further on top, detailed execution phases of each workload are studied, and simpoints corresponding to each phase are added to the portfolio, Obviously the entire suite consists of hundreds of such simpoints; we chose from among them such that a reasonable variety of sizes and apps is covered.
Regarding changing user behavior, further note that since metadata is in memory, static/dynamic binary optimization, similar to BOLT~\cite{bolt} and OCOLOS~\cite{ocolos}, can be applied at runtime to adjust it according to the specific usage scenarios of each user; detailing such mechanism is beyond the scope of this paper.
Table~\ref{tab:benchmarks} provides more information on characteristics of the simpoints in this collection.

\begin{table}[!ht]
    \centering
    \caption{\small Details of the benchmark suite. (CL: cachelines)}
    \resizebox{\columnwidth}{!}{
   \begin{tabular}{|l|l|l|l|l|l|l|l|l|}
    \hline
        ~ & Code & Data & \#instruc- & ~ & Metadata & Metadata & avg. & avg. \#CL \\ 
          & footprint & footprint & tions & \#HBs & size & overhead & \#CL & per HB \\ 
        ~ & (KB) & (KB) & ~ & ~ & (KB) & (\%) & per HB & (last-16) \\ \hline
         App1   & 985.65   & 1964.06   &  13,945,381   &  5,671  & 88.61  & 3.00   & 39.80   & 13.15   \\ \hline
         App2   & 710.58   & 1483.81   &  14,588,360   &  4,079  & 63.73  & 2.90   & 38.15   & 13.21   \\ \hline
         App3   & 626.69   & 1697.19   &  13,260,173   &  4,011  & 62.67  & 2.70   & 38.41   & 13.93   \\ \hline
         App4   & 1678.98   & 5841.69   &  34,543,599   &  11,371  & 177.67  & 2.36   & 38.86   & 13.92   \\ \hline
         App5   & 1043.24   & 4075.38   &  19,804,244   &  6,265  & 97.89  & 1.91   & 42.90   & 13.71   \\ \hline
         App6   & 949.82   & 4514.88   &  19,721,454   &  5,771  & 90.17  & 1.65   & 44.10   & 13.73   \\ \hline
         App7   & 723.34   & 1513.38   &  13,695,360   &  4,112  & 64.25  & 2.87   & 33.67   & 12.91   \\ \hline
         App8   & 832.26   & 2496.44   &  13,395,915   &  4,335  & 67.73  & 2.03   & 29.10   & 12.22   \\ \hline
         App9   & 718.61   & 1947.75   &  14,603,005   &  5,161  & 80.64  & 3.02   & 27.19   & 12.39   \\ \hline
         App10   & 712.57   & 19610.44   &  16,978,094   &  4,858  & 75.91  & 0.37   & 31.10   & 12.99   \\ \hline
         App11   & 231.11   & 939.94   &  12,192,827   &  703  & 10.98  & 0.94   & 41.61   & 13.82   \\ \hline
         App12   & 182.92   & 702.63   &  12,321,070   &  536  & 8.38  & 0.95   & 44.35   & 13.81   \\ \hline
         App13   & 887.07   & 1928.56   &  13,341,744   &  5,755  & 89.92  & 3.19   & 31.71   & 13.47   \\ \hline
         App14   & 918.31   & 2037.25   &  13,166,432   &  6,115  & 95.55  & 3.23   & 30.48   & 13.49   \\ \hline
         App15   & 862.13   & 5703.69   &  32,241,578   &  5,997  & 93.70  & 1.43   & 34.80   & 13.88   \\ \hline
        avg.  & 804.22   &  3,763.8   & 17,186,616   & 4982.67  & 77.85  & 2.17   & 36.41   & 13.37   \\ \hline
    \end{tabular}
    }
    \label{tab:benchmarks}
\end{table}

\begin{table}[t]
\footnotesize
\begin{center}
\caption{\small gem5 Simulation Setup } 
\label{tab:setuptable}
\begin{tabular}{|p{3cm}|p{4cm}|}
\hline
Parameter & Setup \\
\hline
Core & ROB-entries = 512 ;  Issue-Width = 8\\

\hline
Branch prediction & Tournament BP \& Indirect Predictor\\ & BTB Entries = 4096\\ & BTB Tag Size = 16 \\ & RAS Size=32 \\

\hline
L1 I-cache  & 256KB, 8-way \\
L1 D-cache  & 256KB, 8-way  \\
L2 unified cache   & 2MB, 8-way  \\
L2 prefetcher  & Stride Prefetcher\\

\hline
I-TLB size & 64\\
D-TLB size & 64 \\
L2 TLB size & 1280\\

\hline

DRAM & DDR3-1600 x64 \\
\hline

\hline
\end{tabular}
\end{center}
\end{table}

\begin{table}[t]
\footnotesize
\begin{center}
\caption{\small Default settings for DEER evaluation.}
\label{tab:ssrasetup}
\begin{tabular}{|p{3.5cm}|p{3.5cm}|}
\hline
Parameter & Setup \\
\hline

HB metadata format  & SSRA, MLS scheme  \\
Max runahead depth  in SSRA chains & 50 HBs \\
RAS-top prefetch  & Enabled \\
Metadata-cache   & None \\
Metadata load latency  & 400 cycles \\
Prefetching into & Unified L2 cache \\
\hline
Prefetch buffer size & 32 entries\\
Runahead unit RAS size & 16 entries\\
Max \#cachelines to prefetch per HB  & 16  \\

\hline
\end{tabular}
\end{center}
\end{table}

Evaluation is done using gem5 \cite{binkert2011gem5} SE mode for an O3 ARM core, with a two level cache hierarchy consisting of private L1 instruction and data caches and a unified L2 cache, representative of CPUs on most mobile SoCs~\cite{victor}. 
The simulated system setup is summarized in Table~\ref{tab:setuptable}
and remains the same (e.g. the L2-prefetcher) for all techniques.
Commercial mobile processors show a trend of increasing cache sizes~\cite{exynos}, thus we use 256kB L1 caches to model current and future memory systems.

\subsection{Gains over state of the art}
\label{section:gains}

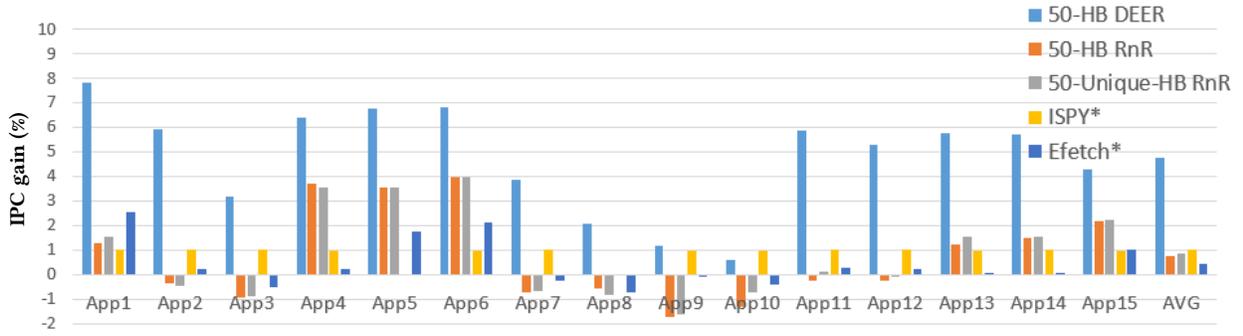
\begin{figure*}[t]
    \centering
    \includegraphics[clip=true, trim=0.7cm 0 0 0, width=0.9\linewidth]{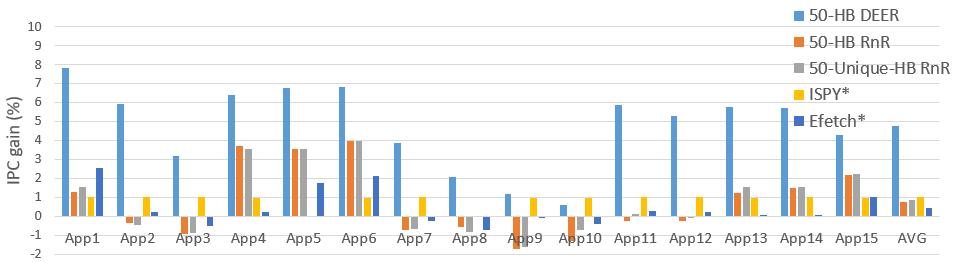}
    \caption{\small IPC gains comparison between DEER and state-of-the-art techniques; higher is better.}
    \setlength{\unitlength}{1cm}
    \begin{picture}(0,0)
        \put(-8.2,2.5){\makebox(0,0)[b]{\rotatebox{90}{\textbf{\small \parbox{7cm} {IPC gain (\%)}}}}}
    \end{picture}
    \label{fig:rivalipcfull}
 \end{figure*}

 \begin{figure*}[t]
    \centering
    \includegraphics[clip=true, trim=0.7cm 0 0 0,  width=0.9\linewidth]{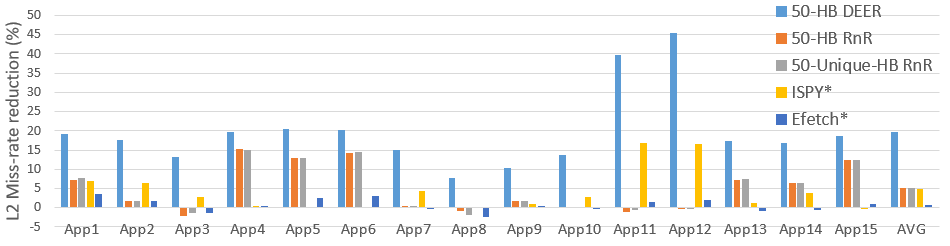}
    \caption{\small L2 I-miss-rate reduction comparison between DEER and state-of-the-art techniques; higher is better.}
    \setlength{\unitlength}{1cm}
    \begin{picture}(0,0)
        \put(-8.2,1.7){\makebox(0,0)[b]{\rotatebox{90}{\textbf{\small \parbox{9cm} {L2 Miss-rate reduction(\%)}}}}}
    \end{picture}
    \label{fig:rivall2missratefull}
 \end{figure*}
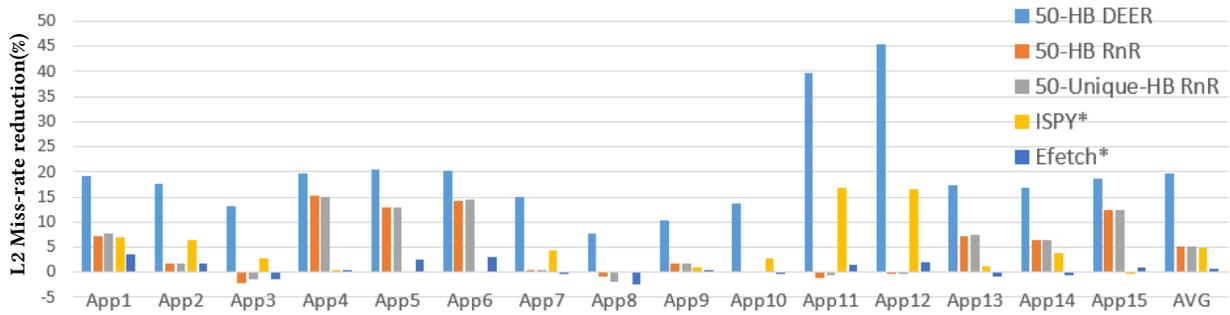

We compare against four rival techniques that represent published state of the art.
Although some of the rivals were originally prefetching to L1 cache in their respective papers, all techniques are fed with the same L2-miss profile and 
are set to prefetch into the L2 cache for consistency and fairness.
Given that L2 misses are the long-latency ones, the bulk of the gain for each technique remains intact while its sensitivity to timeliness is even reduced due to larger size of the L2 cache.

\textbf{RnR (Record-and-Replay)}~\cite{ignite, cache_restore, recap} is a class of techniques that continuously records the accessed addresses, and replays them (i.e. prefetches the address list) at a later time in execution when triggered, e.g. when the first address of the sequence is committed. RnR techniques are able to cover large code footprints with high cache misses, however they require large storage for online recording of the list of addresses. 
We implemented below two variants of RnR for retired instructions to represent the above rivals:
\begin{enumerate}
\item \textbf{50-HB RnR} where we always record the last 50 committed HBs (even if redundant) and assign the list to the oldest HB, so that whenever later that HB is committed again, the list is replayed (prefetched to L2 cache) for it. Obviously this needs a large storage, but we do not put any storage limit on it in our experiments.

\item \textbf{50-Unique-HB RnR}  This represents Hierarchical Prefetching ~\cite{zhang2025hierarchical} but using Hyperblocks as a coarse grain measure for recording, and thus called Hierarchical-Prefetching* to note the difference. 
Each time a trigger HB is seen, we record the trace of the next 50 \textit{unique} HBs that appear after, and replay them (i.e. prefetch to L2 cache) upon observing the trigger HB again. No storage limitation is applied, and no instruction overhead is accounted for.

\end{enumerate}

\textbf{I-Spy*:} We produced the list of PC locations (\emph{trigger-PC}) for prefetch-insertion per missed PC (\emph{target-PC}) using I-Spy~\cite{ispy} code repository (this needed us to produce the required inputs to their tools, from the ARM instruction traces that we had obtained from our benchmarks); then our gem5 model prefetched the \emph{target-PC} whenever the designated \emph{trigger-PC} was retired. Note that this does not model the instruction overhead associated with original I-Spy, and hence we call this I-Spy*.

\textbf{EFetch*:} EFetch~\cite{efetch} decides the code region to prefetch based on a software hint that tells what action the user took; it also prefetches large functions piece-by-piece for timeliness. 
There is no \emph{user hint} in our workloads to simulate here, but we implemented same piece-by-piece prefetching of function codes as in EFetch, and we did that for same 1-level call depth that ~\cite{efetch} decided as their reasonable choice; we call it EFetch* to clarify the difference.

For more explanation on I-Spy and EFetch, see Section~\ref{section:codesigns}.

\textbf{50-HB DEER:} Table ~\ref{tab:ssrasetup} shows the settings used for DEER when comparing against rivals. Results of design-space exploration for those parameters are given in Section~\ref{section:dse} as well as sensitivity analyses below.

\subsubsection{Performance Gains}

Figures~\ref{fig:rivalipcfull} and~\ref{fig:rivall2missratefull} show the IPC gains and L2 I-miss-rate reductions for each evalutated mechanism respectively.
On our mobile workloads, 50-HB DEER reduces L2 I-miss-rate by an average of 19.9\%, while 50-HB RnR, 50-Unique-HB RnR, and I-Spy* achieve 5.02\%, 5.08\%, and 4.8\% respectively, and EFetch* gets only 0.7\% improvement.
This can be explained by below advantages of DEER over the rivals:
(i) prefetching for multiple cachelines over a deeper runahead and call-stack, unlike EFetch* which goes only one level deep in the call stack;
(ii) skipping over loops and recursions, and prefetching for the return paths, unlike I-Spy*, and 50 HB RnR.
(iii) using the most-likely path prediction instead of recording the last observed path as RnR methods do, which achieves better performance in programs with less {\it immediately-repeating} behavior. 
To further validate this last point, we plotted the prediction-accuracy metric---see Section~\ref{motivation}---for the 50-Unique-HB RnR method; Figure \ref{fig:accuracyRnR} shows the drop in accuracy compared to DEER and explains the inferior gains. 

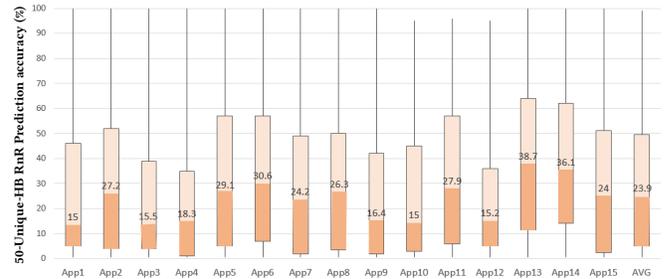
\begin{figure}
    \centering
    \includegraphics[width=1\linewidth]{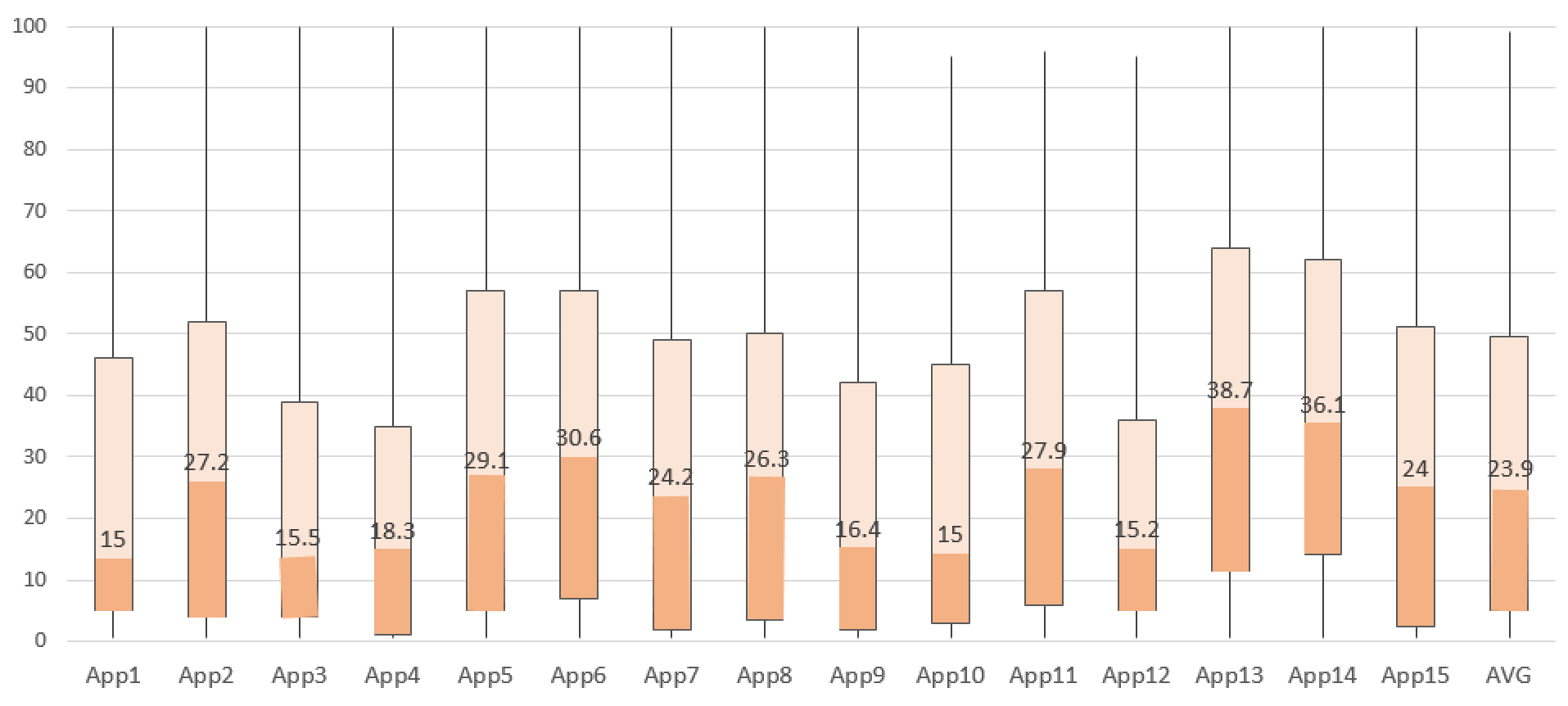}
    \caption{\small Path prediction accuracy of 50-Unique-HB RnR (Hierarchical-Prefetching*), in a box-and-whisker plot.}
     \setlength{\unitlength}{1cm}
    \begin{picture}(0,0)
        \put(-4.3,1.7){\makebox(0,0)[b]{\rotatebox{90}{\textbf{\tiny \parbox{5cm} {50-Unique-HB RnR Prediction accuracy (\%)}}}}}
        
    \end{picture}
    \label{fig:accuracyRnR}
 \end{figure}

DEER essentially predicts the most likely future path based on the examined profile, and tries to prefetch the corresponding cachelines; thus it helps to cover all 3 classes of misses. One specific case is on App11 and App12 that are smaller than the L1-I (see Table~\ref{tab:benchmarks}) and hence, they suffer only from cold misses; DEER has been successful in covering most of these cold misses---cf. the oracle in Figure~\ref{fig:oracle}. This is important because our study of task scheduling on mobile phones reveals that context-switches are happening extremely frequently, even every fraction of million instructions; consequently, the core severely suffers from 
cold misses, where DEER proves successful in covering them.

  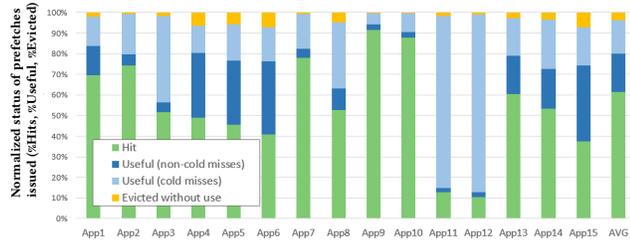
\begin{figure}[t]
    \hspace{0.4cm}
    \includegraphics[clip=true, trim=0.2cm 0 0 0, width=0.94\linewidth]{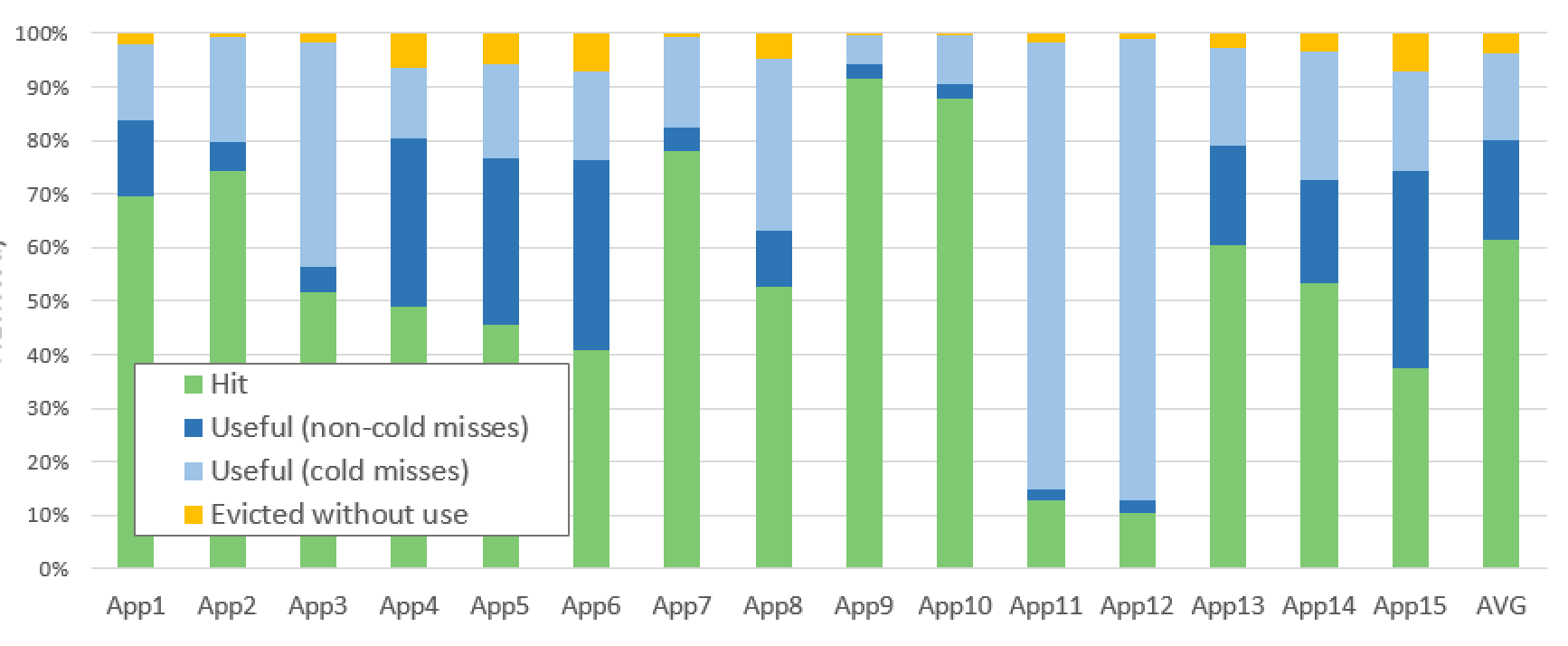}
    \caption{\small Breakdown of  prefetches issued by \textit{DEER} for each simpoint; Status can be \textit{hit} (already in L2), \textit{useful} (accessed by the main thread) broken down further to covering \textit{cold} vs. \textit{non-cold} misses, and \textit{evicted without use} (wrong/not-timely).}
    \label{fig:usefulness}
    \setlength{\unitlength}{1cm}
    \begin{picture}(0,0)
        \put(-4,2.8){\makebox(0,0)[b]{\rotatebox{90}{\textbf{\tiny \parbox{7cm} {Normalized status of  prefetches \\ issued (\%Hits, \%Useful, \%Evicted)}}}}}
 
    \end{picture}

  \end{figure}

\subsubsection{Prefetch Usefulness}
\label{sec:prfusefullness}

We already evaluated and discussed the {\it accuracy} of the path predicted by our MLS scheme in Figure~\ref{fig:accuracy}.
To complement it with a {\it usefulness} study, we provide the breakdown of the issued prefetches into three categories: 
(i) \emph{hit:} are prefetches that turn out to be already in the L2 cache and are essentially "redundant",
(ii) \emph{useful:} prefetches that are actually accessed after filled in the L2 cache by DEER
(this class is further broken down to covering \textit{cold} vs. \textit{capacity/conflict} misses based on coldness of the prefetched cacheline),
and (iii) \emph{evicted without use:} which are wasteful and polluting the cache since they did not help and also evicted a potentially useful cacheline; note how the last class can be either a {\it wrong} prefetch or {\it not timely}.
Figure~\ref{fig:usefulness} shows the above breakdown; it also confirms: 
1) the positive relationship between usefulness of the prefetches and the gains in Figure~\ref{fig:rivalipcfull};
2) both cold and capacity/conflict misses are important and equally contributing on average,
3) on smaller simpoints, e.g. Apps 11 and 12, the gain mostly comes from covering cold misses, whereas on others, the share of capacity/conflict misses can be higher depending on the execution and memory-access patterns.
The experiments show that on average, the correction mechanism occurs on 28.8\% of the predicted paths; however note that this ratio is not the best measure of success/failure since the predicted and actual paths may still re-converge later. Consequently, we rely on the IOU measure, depicted on Figure~\ref{fig:accuracy}, as a better measure of accuracy for the prefetches.

\begin{table}[t]
\footnotesize
\begin{center}
\caption{\small Effective runahead depth.} %
\label{tab:depth}
\resizebox{!}{2cm}{
    \begin{tabular}{|p{1cm}|p{1.8cm}|p{1.8cm}|p{1.8cm}|}
    \hline
    Trace  & Dynamic depth (\#Insts.) & Static depth (\#Insts.) &  \#Cycles skipped \\
    \hline
    APP1 & 796 & 378 &	2.1 \\
    APP2 & 555 &	390 &	1.4  \\
    APP3 & 553 &	399 &	1.4 \\
    APP4 & 9562 &	1061 &	9.0 \\
    APP5 & 738	& 341	& 2.2\\
    APP6 & 811	& 379	& 2.1 \\
    APP7 & 3265 &	465	& 7.0\\
    APP8 & 1257 &	326	& 3.8 \\
    APP9 & 409	 & 285	 & 1.4 \\
    APP10 & 444	 & 311 &	1.4\\
    APP11 & 25509 &	768	& 33.2 \\
    APP12 & 27745 &	874	& 31.7 \\
    APP13 & 998	& 397 & 2.5 \\
    APP14 & 667	& 316 &	2.1 \\
    APP15 & 1150 &	358	& 3.2 \\
    
    \hline
    
    \end{tabular}
    }
\end{center}
\end{table}

\subsubsection{Contributions of DEER hardware features}
We described two features in Section~\ref{section:hardware}: {\it prefetch-on-refill} and {\it RAS-top-prefetch}.
Experiments show that compared to an unlimited metadata-cache case, the prefetch-on-refill feature 
manages to revive 98.8\% of the performance gains.
This proves that the runahead is deep enough ahead of fetch such that despite the long metadata-access latency, the corresponding I-cacheline prefetches are still timely. 

Algorithm~\ref{algo:ssra} shows two metadata lines are loaded per invocation: one for the trigger-PC, and another one for the RAS-top-PC. 
Figure~\ref{fig:contributions} shows the individual and combined contributions of the two metadata lines. 
The RAS-top one contributes slightly less than the trigger-PC one when applied alone, but when combined, on average 9.6\% additional IPC gain is obtained: average IPC improvement increases to 4.78\% from 4.36\%.

\begin{figure}[t]
    \centering
    \includegraphics[width=1\linewidth]{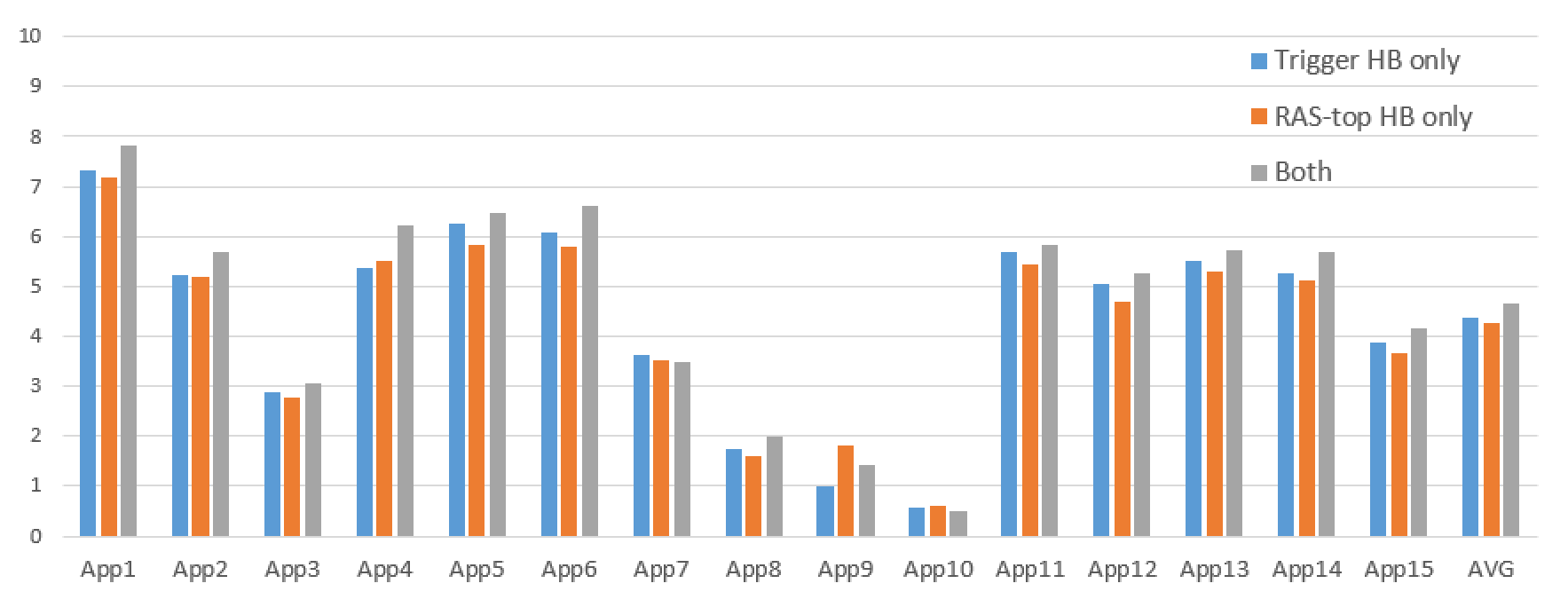}
    \caption{\small DEER IPC gains (\%) when prefetching for (i) Trigger HB only (ii) RAS-top HB only (iii) both HBs.}
     \setlength{\unitlength}{1cm}
    \begin{picture}(0,0)
        \put(-4.3,1.8){\makebox(0,0)[b]{\rotatebox{90}{\textbf{\tiny \parbox{5cm} {IPC gain (\%) for different prefetch\\modes}}}}}
        
    \end{picture}
    \label{fig:contributions}
 \end{figure}
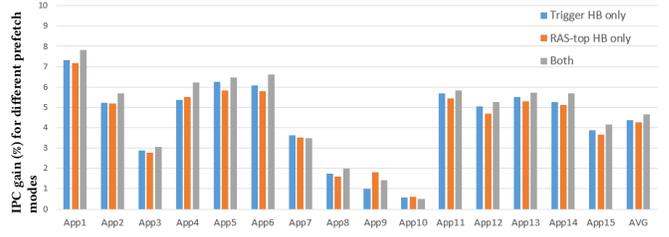

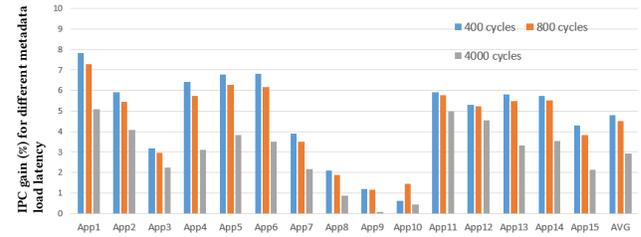
\begin{figure}[t]
    \centering
    \includegraphics[clip=true, trim=2.5cm 0 0 0, width=0.94\linewidth]{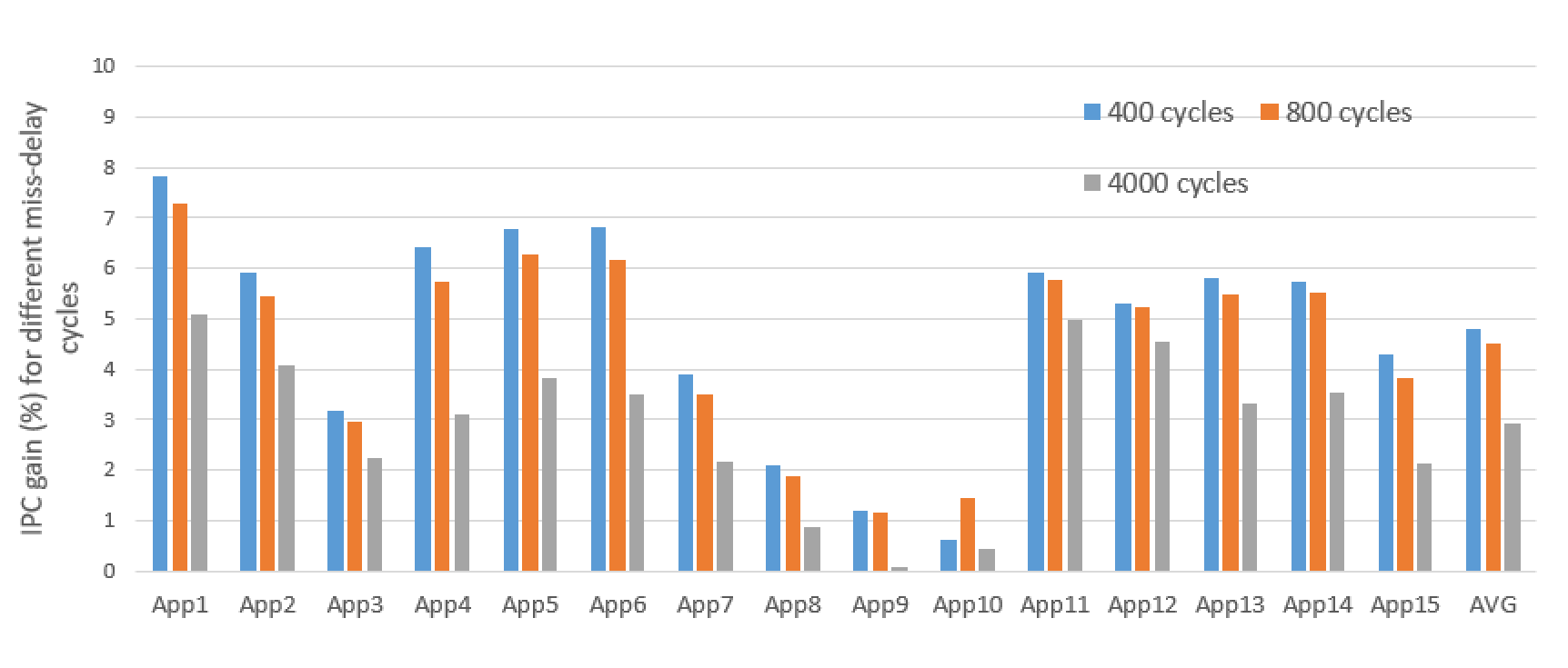}
    \caption{\small Sensitivity of IPC gain (\%) to metadata-load latency.}
    \setlength{\unitlength}{1cm}
    \begin{picture}(0,0)
        \put(-4.2 ,1.6){\makebox(0,0)[b]{\rotatebox{90}{\textbf{\tiny \parbox{5cm} {IPC gain (\%) for different metadata \\load latency}}}}}
    \end{picture}
    \label{fig:hbcache-delay-ipc}
 \end{figure}

\subsubsection{Effective runahead depth} 
\label{effective-runahead}
Table ~\ref{tab:depth} shows the {\it effective runahead depth}, defined as the number of instructions (dynamic and static in two columns) in $S_{PC}^{Ex}$ when path prediction accuracy is above 60\%---see Section~\ref{motivation} for definitions of $S_{PC}^{Ex}$ and prediction accuracy.
Last column of the table gives the ratio between number of dynamic vs. static instructions; i.e. average number of cycles skipped by the runahead.
The median number of cycles  skipped is 2.2, but the extreme cases have up to 15x more cycles; this shows the significance of the cycle-skipping feature in allowing to go deeper beyond loops.
We further observed that RAS-top-prefetch provides a similar but slightly smaller depth. 
It is worth noting how this $N=$ \textit{effective static runahead depth} is the static-timing mechanism that DEER employs to time the prefetching; At each trigger PC, DEER prefetches for the upcoming $N$ instructions where $N$ is decided by the number of HBs in the runahead chain. We have experimentally decided that 50-HBs-ahead is a good compromise between timeliness and cache pollution; the advantage is hardware simplicity at the cost of some lost gains. Nonetheless, a dynamically adaptive runahead-depth adjuster with a control feedback loop fed with usefulness-/eviction-rate of injected prefetches would reap even more gains.

\subsection{Sensitivity Analysis}
\subsubsection{Metadata-load latency}

The metadata table is put in a non-cacheable area in main memory to avoid impacts on cache memories when a metadata line is requested by the core causing considerable request latency.
Figure~\ref{fig:hbcache-delay-ipc} illustrates the loss of gain coming from different delays which lowers at higher load-latencies. Yet, we see that the runahead is deep enough that many I-cacheline prefetches are still timely.
The slight anomaly on App10 is explained by high redundant prefetches (Figure~\ref{fig:usefulness}) and the randomness caused by having more metadata lines in flight simultaneously at higher access latencies.

\subsubsection{Max runahead depth}
To examine the impact of the runahead depth in as isolated as possible, we employ no metadata miss and the dynamic runahead scheme, instead of the SSRA. 

Results show that (if not limited by the static depth of the SSRA) we could achieve better gains with increasing the runahead until a certain point defined by the trade-off between accuracy and timeliness. 
Hence, for most workloads, after the depth of 150 HBs, gains begin to drop.

\begin{figure}[t]
    \centering
    \includegraphics[width=0.95\linewidth]{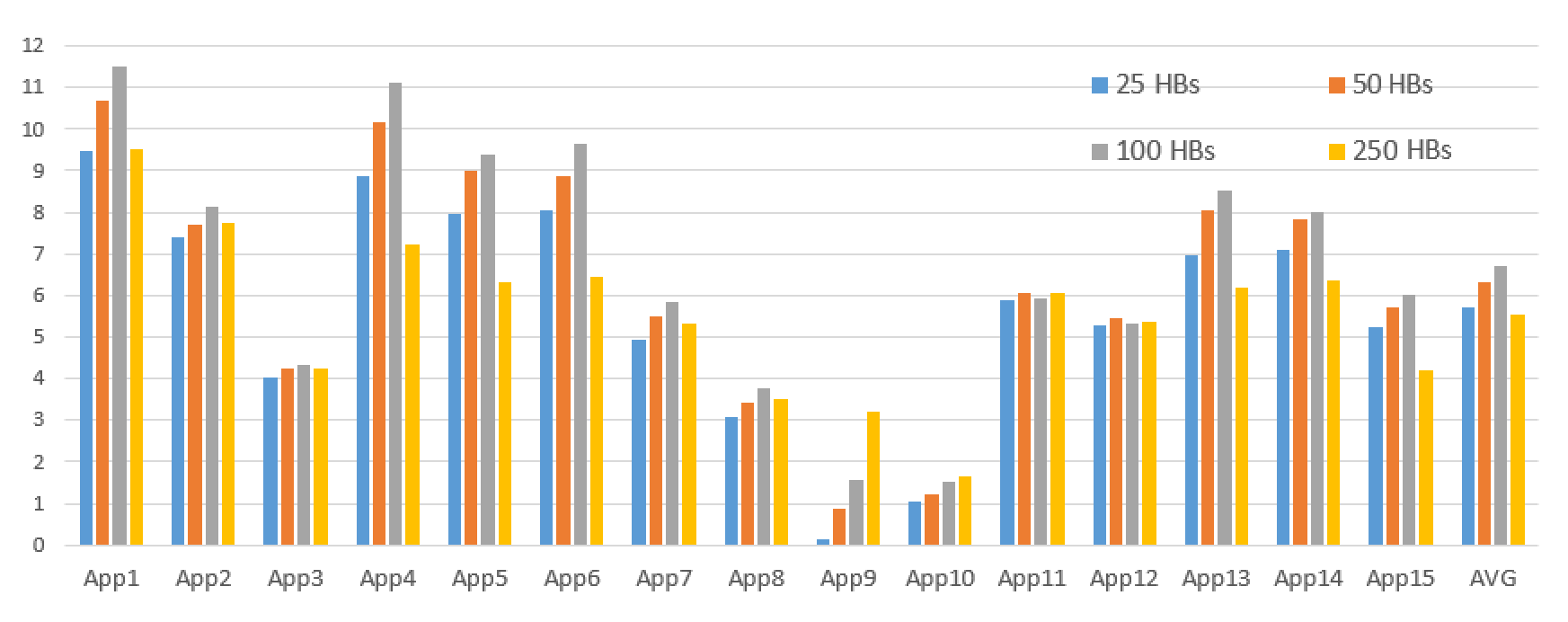}
    \caption{\small Sensitivity of IPC gains to the chosen maximum runahead depth in an ideal dynamic runahead scheme.}
    \setlength{\unitlength}{1cm}
    \begin{picture}(0,0)
        \put(-4.2,1.7){\makebox(0,0)[b]{\rotatebox{90}{\textbf{\tiny  \parbox{6cm} {IPC gain (\%) for different  runahead\\depths}}}}}
    \end{picture}
    
    \label{fig:diffra-ipc}
 \end{figure}
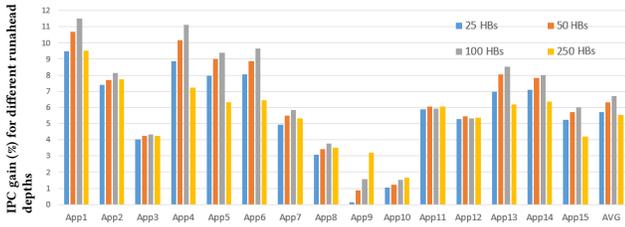

\subsection{Practical Values for Design Parameters}
\label{section:dse}
Figure~\ref{fig:hw} shows major storage elements of DEER hardware component in blue. In this section we discuss the reasons behind our choices of sizes and values for them.

\subsubsection{Number of cachelines per metadata line}

Table~\ref{tab:benchmarks} shows that each SSRA metadata line contains roughly 36 cachelines on average. We examined the impact of reducing that number so as to reduce the overhead of metadata storage. When trying lower number of cachelines per metadata line, we pick the last N cachelines in the SSRA chain; i.e. the cachelines corresponding to the farthest point in future in the predicted execution flow; this gives more chance for a timely prefetch.
Based on the results shown in Figures~\ref{fig:lastn-ipc}, we chose to target 16 cachelines per HB metadata as a reasonable compromise between gains and storage overhead. Then we examined the proximity of the chosen cachelines as well as dispersity of those neighbourhood regions in memory; accordingly we chose 512 Byte regions as the granule that gives reasonable utilization of the bitmaps, and also covered six such regions in the metadata encoding as described in Section~\ref{sec:sw-hw-interface}. 

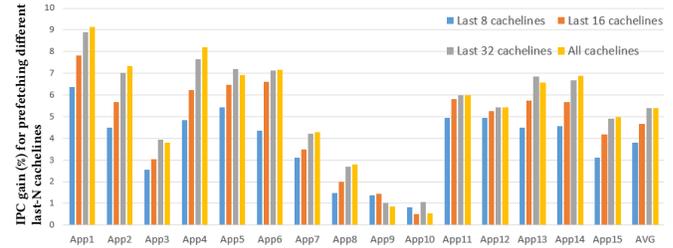
\begin{figure}[t]
    \centering
    \includegraphics[clip=true, trim=1.2cm 0 0 0, width=1\linewidth]{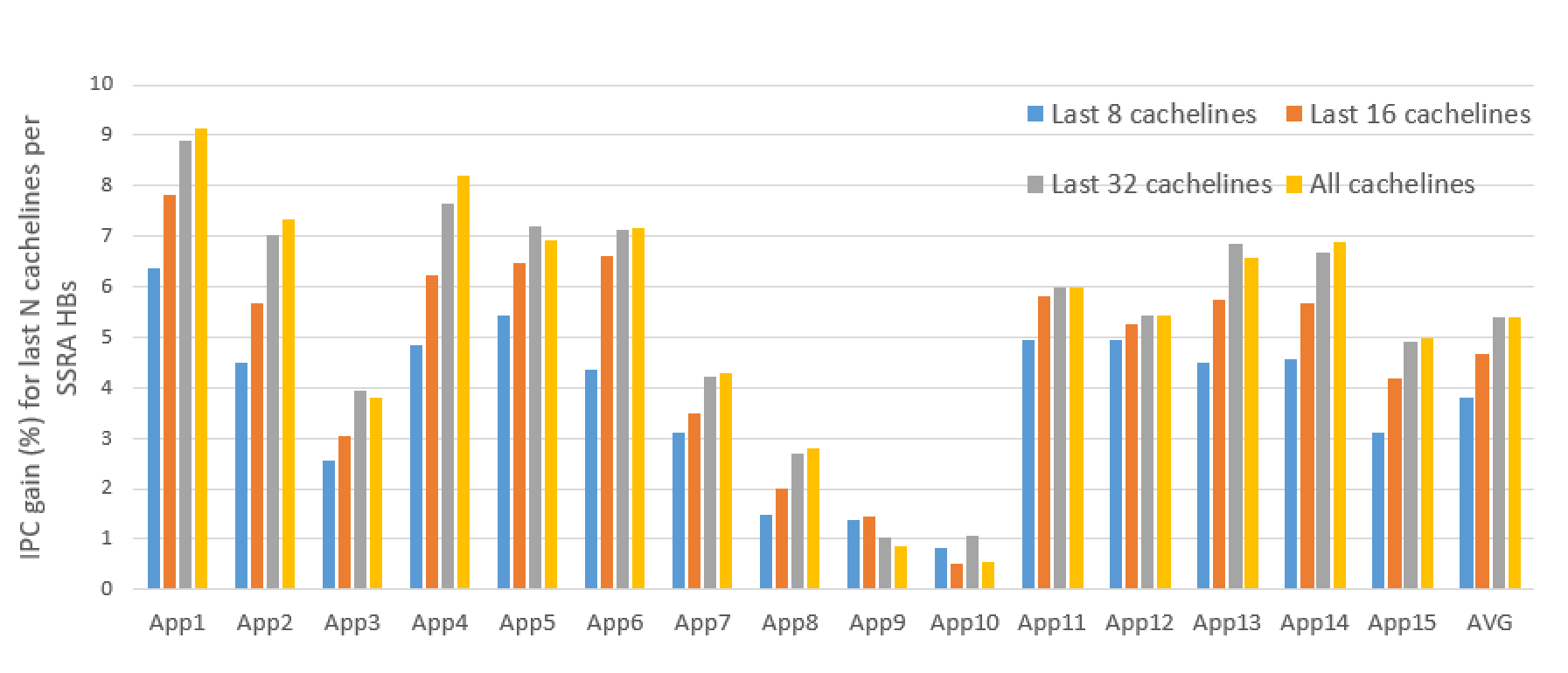}
    \caption{\small DEER IPC gains for different last-N cachelines prefetched for each SSRA chain.}

    \setlength{\unitlength}{1cm}
    \begin{picture}(0,0)
        \put(-4.4,1.9){\makebox(0,0)[b]{\rotatebox{90}{\textbf{\tiny  \parbox{6cm} {IPC gain (\%) for prefetching different \\ last-N cachelines}}}}}
    \end{picture}
    
    \label{fig:lastn-ipc}
 \end{figure}

\subsubsection{Prefetch buffer size}

Figure~\ref{fig:qlimit-ipc} shows that in smaller buffer sizes such as 8 and 16 entries, gains reduce since this tight limit leads to the drop of useful prefetches. 
App9 and App10 behave differently; this can be explained by the high redundant prefetches in them---see Figure~\ref{fig:usefulness}: the imposed tighter limit reduces the pressure on the LSU resources by redundant prefetches. 
We further observe that at 32 entries, we get maximum gain that remains almost constant at higher sizes, until a point where timeliness becomes an issue due to favoring late prefetches to more recent ones. 

 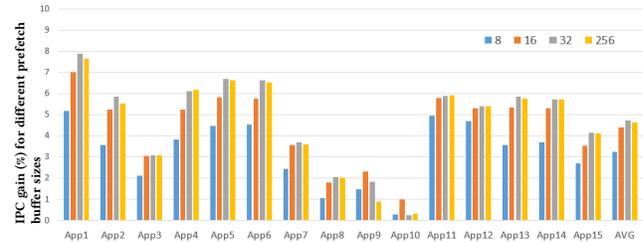
\begin{figure}[t]
    \centering
    \includegraphics[clip=true, trim=1.5cm 0 0 0, width=0.97\linewidth]{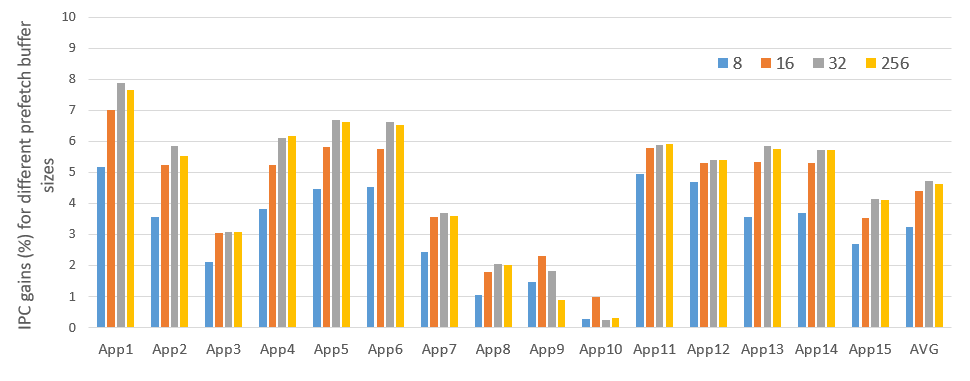}
    \caption{\small DEER IPC gains vs. prefetch buffer size.}
    
    \setlength{\unitlength}{1cm}
    \begin{picture}(0,0)
        \put(-4.2,1.5){\makebox(0,0)[b]{\rotatebox{90}{\textbf{\tiny  \parbox{6cm} {IPC gain (\%) for different prefetch \\ buffer sizes}}}}}
    \end{picture}
    \label{fig:qlimit-ipc}
 \end{figure}

\subsubsection{RAS size}
On average, unlimited RAS size gives 4.78\% IPC gain on our simpoints, which remains almost constant until the RAS size of 16 entries; then it reduces to 4.68\% for 8 entries.
Unlike the dynamic runahead mechanism that employs RAS intensively for each runahead, the SSRA method only uses RAS for RAS-top-prefetch. This justifies the observed low sensitivity of the RAS-top-prefetch gains to RAS size.

\subsubsection{Overheads}
\label{overheads}
DEER overheads can be divided into three classes:
\begin{enumerate}
    \item {\it Storage on chip}: this includes the prefetch-buffer, RAS, and fetched-metadata buffer, which are respectively 32 entries (32x6 bytes), 16 entries (16x6 bytes), and a single entry (16 bytes), to a tally of only 304 bytes.
    The remainder of the DRU is trivial glue logic and thus, small compared to the storage elements.
    \item {\it Metadata size on the binary}: This is marginal compared to the size of the program binaries---see Section~\ref{sec:sw-hw-interface}.
    \item {\it Metadata storage on memory at run time}: Employing hash scheme such as cuckoo~\cite{pagh2004cuckoo} and Murmur3 hashing functions \cite{murmur} typically doubles the memory footprint of the metadata, which is still marginal compared to the size of the binaries and libraries loaded into the memory at run-time.
\end{enumerate}
\section{Summary and Conclusion}
We showed a simple profile-based path predictor can fairly well foretell the upcoming cache lines as the execution proceeds, and thus a corresponding co-designed prefetcher can effectively prefetch them in advance. 
We use this predictor to continuously and timely refill a smaller cache (and indeed any other hardware table) such that it effectively provides same performance as a much larger one lacking the feature. 
This is important because in the extremely-frequent-preemption environment of mobile workloads, that are run on big-little SoCs with complex cache hierarchies, and where severe cross-library calls with very deep call stacks are the norm, conventional software-only or hardware-only techniques either fall short in accuracy, coverage, and timeliness, or are too expensive. 

We provided a SW/HW collaboration scheme that realized a low cost Kilo-instructions deep prefetch mechanism for instructions and showed that it works effectively in heavily frontend bound mobile workloads by eliminating most cold as well as capacity and conflict misses.
DEER is a static most-likely-path predictor that continuously corrects itself based on the retired control flow.
We devised a light weight SW/HW interface that requires setting only one system register to point to the metadata table in memory; this register is additionally saved/restored upon context switch.
DEER works on binaries without needing source codes of the libraries.
By relieving the hardware from path-prediction chores, DEER eliminates the need for area-/power-expensive on-chip memories to store metadata, and accordingly beats full-hardware rivals by providing over 4x higher gains at two orders of magnitude lower cost.

\bibliographystyle{plain}
\bibliography{references}

\begin{thebibliography}{10}

\bibitem{arm2020}
Arm architecture 2020 extensions.
\newblock In {\em
  https://static.linaro.org/connect/lvc20/presentations/LVC20-214-0.pdf}.

\bibitem{armbiglittle}
Arm big.little technology.
\newblock In {\em https://www.arm.com/en/technologies/big-little}.

\bibitem{armcmc1}
Arm ltd. arm cortex-a78ae core technical reference manual revision r0p1,
  cpuectlr el1, cpu extended control register, el1.

\bibitem{armcmc2}
Arm ltd. arm cortex-x2 core technical reference manual r2p0, imp cpuectlr el1,
  cpu extended control register.

\bibitem{armarch}
Development of the arm architecture.
\newblock In {\em
  https://developer.arm.com/documentation/102404/0201/Development-of-the-Arm-architecture}.

\bibitem{Exynos5}
Exynos 5 octa.
\newblock In {\em http://www.samsung.com/global/business/
  semiconductor/file/media/Exynos 5 Octa.pdf}.

\bibitem{geekbench}
Geekbenc home page.
\newblock In {\em https://www.primatelabs.com/}.

\bibitem{lbr}
An introduction to last branch records.
\newblock In {\em https://lwn.net/Articles/680985/}.

\bibitem{brbe}
Lucas prates. 2020. add support for the branch record buffer extention.
\newblock In {\em https://reviews.llvm.org/D92389}.

\bibitem{murmur}
Murmur3 hash.
\newblock In {\em https://github.com/aappleby/smhasher/wiki/MurmurHash3}.

\bibitem{Qualcomm808}
Qualcomm snapdragon 808 processor.
\newblock In {\em
  https://www.qualcomm.com/products/mobile/snapdragon/smartphones/snapdragon-8-series-mobile-platforms/snapdragon-processors-808}.

\bibitem{spec}
Spec home page.
\newblock In {\em https:https://www.spec.org/}.

\bibitem{ainsworth2024triangel}
Sam Ainsworth and Lev Mukhanov.
\newblock Triangel: A high-performance, accurate, timely on-chip temporal
  prefetcher.
\newblock {\em arXiv preprint arXiv:2406.10627}, 2024.

\bibitem{ansari2022mana}
Ali Ansari, Fatemeh Golshan, Rahil Barati, Pejman Lotfi-Kamran, and Hamid
  Sarbazi-Azad.
\newblock Mana: Microarchitecting a temporal instruction prefetcher.
\newblock {\em IEEE Transactions on Computers}, 72(3):732--743, 2022.

\bibitem{ansari2020divide}
Ali Ansari, Pejman Lotfi-Kamran, and Hamid Sarbazi-Azad.
\newblock Divide and conquer frontend bottleneck.
\newblock In {\em 2020 ACM/IEEE 47th Annual International Symposium on Computer
  Architecture (ISCA)}, pages 65--78. IEEE, 2020.

\bibitem{ayers2018memory}
Grant Ayers, Jung~Ho Ahn, Christos Kozyrakis, and Parthasarathy Ranganathan.
\newblock Memory hierarchy for web search.
\newblock In {\em 2018 IEEE International Symposium on High Performance
  Computer Architecture (HPCA)}, pages 643--656. IEEE, 2018.

\bibitem{asmdb}
Grant Ayers, Nayana~Prasad Nagendra, David~I. August, Hyoun~Kyu Cho, Svilen
  Kanev, Christos Kozyrakis, Trivikram Krishnamurthy, Heiner Litz, Tipp
  Moseley, and Parthasarathy Ranganathan.
\newblock Asmdb: understanding and mitigating front-end stalls in
  warehouse-scale computers.
\newblock In {\em Proceedings of the 46th International Symposium on Computer
  Architecture}, ISCA '19, page 462–473, New York, NY, USA, 2019. Association
  for Computing Machinery.

\bibitem{binkert2011gem5}
Nathan Binkert, Bradford Beckmann, Gabriel Black, Steven~K Reinhardt, Ali
  Saidi, Arkaprava Basu, Joel Hestness, Derek~R Hower, Tushar Krishna, Somayeh
  Sardashti, et~al.
\newblock The gem5 simulator.
\newblock {\em ACM SIGARCH computer architecture news}, 39(2):1--7, 2011.

\bibitem{chacon2023characterization}
Gino Chacon, Nathan Gober, Krishnendra Nathella, Paul~V Gratz, and Daniel~A
  Jim{\'e}nez.
\newblock A characterization of the effects of software instruction prefetching
  on an aggressive front-end.
\newblock In {\em 2023 IEEE International Symposium on Performance Analysis of
  Systems and Software (ISPASS)}, pages 61--70. IEEE, 2023.

\bibitem{efetch}
Gaurav Chadha, Scott Mahlke, and Satish Narayanasamy.
\newblock Efetch: optimizing instruction fetch for event-driven
  webapplications.
\newblock In {\em Proceedings of the 23rd International Conference on Parallel
  Architectures and Compilation}, PACT '14, page 75–86, New York, NY, USA,
  2014. Association for Computing Machinery.

\bibitem{cache_restore}
David Daly and Harold~W. Cain.
\newblock Cache restoration for highly partitioned virtualized systems.
\newblock In {\em IEEE International Symposium on High-Performance Comp
  Architecture}, pages 1--10, 2012.

\bibitem{doweck2017inside}
Jack Doweck, Wen-Fu Kao, Allen Kuan-yu Lu, Julius Mandelblat, Anirudha
  Rahatekar, Lihu Rappoport, Efraim Rotem, Ahmad Yasin, and Adi Yoaz.
\newblock Inside 6th-generation intel core: New microarchitecture code-named
  skylake.
\newblock {\em IEEE Micro}, 37(2):52--62, 2017.

\bibitem{ferdman2012clearing}
Michael Ferdman, Almutaz Adileh, Onur Kocberber, Stavros Volos, Mohammad
  Alisafaee, Djordje Jevdjic, Cansu Kaynak, Adrian~Daniel Popescu, Anastasia
  Ailamaki, and Babak Falsafi.
\newblock Clearing the clouds: a study of emerging scale-out workloads on
  modern hardware.
\newblock {\em Acm sigplan notices}, 47(4):37--48, 2012.

\bibitem{ferdman2011proactive}
Michael Ferdman, Cansu Kaynak, and Babak Falsafi.
\newblock Proactive instruction fetch.
\newblock In {\em Proceedings of the 44th Annual IEEE/ACM International
  Symposium on Microarchitecture}, pages 152--162, 2011.

\bibitem{ferdman2008temporal}
Michael Ferdman, Thomas~F Wenisch, Anastasia Ailamaki, Babak Falsafi, and
  Andreas Moshovos.
\newblock Temporal instruction fetch streaming.
\newblock In {\em 2008 41st IEEE/ACM International Symposium on
  Microarchitecture}, pages 1--10. IEEE, 2008.

\bibitem{intel-lunar-lake}
Arik Gihon.
\newblock { Lunar Lake Architecture Session }.
\newblock In {\em 2024 IEEE Hot Chips 36 Symposium (HCS)}, pages 1--49, Los
  Alamitos, CA, USA, August 2024. IEEE Computer Society.

\bibitem{godala2024pdip}
Bhargav~Reddy Godala, Sankara~Prasad Ramesh, Gilles~A Pokam, Jared Stark, Andre
  Seznec, Dean Tullsen, and David~I August.
\newblock Pdip: Priority directed instruction prefetching.
\newblock In {\em Proceedings of the 29th ACM International Conference on
  Architectural Support for Programming Languages and Operating Systems, Volume
  2}, pages 846--861, 2024.

\bibitem{exynos}
Brian Grayson, Jeff Rupley, Gerald~Zuraski Zuraski, Eric Quinnell, Daniel~A.
  Jiménez, Tarun Nakra, Paul Kitchin, Ryan Hensley, Edward Brekelbaum, Vikas
  Sinha, and Ankit Ghiya.
\newblock Evolution of the samsung exynos cpu microarchitecture.
\newblock In {\em 2020 ACM/IEEE 47th Annual International Symposium on Computer
  Architecture (ISCA)}, pages 40--51, 2020.

\bibitem{ishii2021re}
Yasuo Ishii, Jaekyu Lee, Krishnendra Nathella, and Dam Sunwoo.
\newblock Re-establishing fetch-directed instruction prefetching: An industry
  perspective.
\newblock In {\em 2021 IEEE International Symposium on Performance Analysis of
  Systems and Software (ISPASS)}, pages 172--182. IEEE, 2021.

\bibitem{jamilan2022apt}
Saba Jamilan, Tanvir~Ahmed Khan, Grant Ayers, Baris Kasikci, and Heiner Litz.
\newblock Apt-get: Profile-guided timely software prefetching.
\newblock In {\em Proceedings of the Seventeenth European Conference on
  Computer Systems}, pages 747--764, 2022.

\bibitem{kanev2015profiling}
Svilen Kanev, Juan~Pablo Darago, Kim Hazelwood, Parthasarathy Ranganathan, Tipp
  Moseley, Gu-Yeon Wei, and David Brooks.
\newblock Profiling a warehouse-scale computer.
\newblock In {\em Proceedings of the 42nd annual international symposium on
  computer architecture}, pages 158--169, 2015.

\bibitem{victor}
Victor Kariofillis and Natalie~Enright Jerger.
\newblock Workload characterization of commercial mobile benchmark suites.
\newblock In {\em 2024 IEEE International Symposium on Performance Analysis of
  Systems and Software (ISPASS)}, pages 73--84, 2024.

\bibitem{kaynak2013shift}
Cansu Kaynak, Boris Grot, and Babak Falsafi.
\newblock Shift: Shared history instruction fetch for lean-core server
  processors.
\newblock In {\em Proceedings of the 46th Annual IEEE/ACM International
  Symposium on Microarchitecture}, pages 272--283, 2013.

\bibitem{kaynak2015confluence}
Cansu Kaynak, Boris Grot, and Babak Falsafi.
\newblock Confluence: unified instruction supply for scale-out servers.
\newblock In {\em Proceedings of the 48th International Symposium on
  Microarchitecture}, pages 166--177, 2015.

\bibitem{khan2021twig}
Tanvir~Ahmed Khan, Nathan Brown, Akshitha Sriraman, Niranjan~K Soundararajan,
  Rakesh Kumar, Joseph Devietti, Sreenivas Subramoney, Gilles~A Pokam, Heiner
  Litz, and Baris Kasikci.
\newblock Twig: Profile-guided btb prefetching for data center applications.
\newblock In {\em MICRO-54: 54th Annual IEEE/ACM International Symposium on
  Microarchitecture}, pages 816--829, 2021.

\bibitem{ispy}
Tanvir~Ahmed Khan, Akshitha Sriraman, Joseph Devietti, Gilles Pokam, Heiner
  Litz, and Baris Kasikci.
\newblock I-spy: Context-driven conditional instruction prefetching with
  coalescing.
\newblock In {\em 2020 53rd Annual IEEE/ACM International Symposium on
  Microarchitecture (MICRO)}, pages 146--159, 2020.

\bibitem{kolli2013rdip}
Aasheesh Kolli, Ali Saidi, and Thomas~F Wenisch.
\newblock Rdip: Return-address-stack directed instruction prefetching.
\newblock In {\em Proceedings of the 46th Annual IEEE/ACM International
  Symposium on Microarchitecture}, pages 260--271, 2013.

\bibitem{kumar2017boomerang}
Rakesh Kumar, Cheng-Chieh Huang, Boris Grot, and Vijay Nagarajan.
\newblock Boomerang: A metadata-free architecture for control flow delivery.
\newblock In {\em 2017 IEEE International Symposium on High Performance
  Computer Architecture (HPCA)}, pages 493--504. IEEE, 2017.

\bibitem{coopprfm}
Chi-Keung Luk and T.C. Mowry.
\newblock Cooperative prefetching: compiler and hardware support for effective
  instruction prefetching in modern processors.
\newblock In {\em Proceedings. 31st Annual ACM/IEEE International Symposium on
  Microarchitecture}, pages 182--193, 1998.

\bibitem{luk1998cooperative}
Chi-Keung Luk and Todd~C Mowry.
\newblock Cooperative prefetching: Compiler and hardware support for effective
  instruction prefetching in modern processors.
\newblock In {\em Proceedings. 31st Annual ACM/IEEE International Symposium on
  Microarchitecture}, pages 182--193. IEEE, 1998.

\bibitem{nesbit2004data}
Kyle~J Nesbit and James~E Smith.
\newblock Data cache prefetching using a global history buffer.
\newblock In {\em 10th International Symposium on High Performance Computer
  Architecture (HPCA'04)}, pages 96--96. IEEE, 2004.

\bibitem{oh2024udp}
Surim Oh, Mingsheng Xu, Tanvir~Ahmed Khan, Baris Kasikci, and Heiner Litz.
\newblock Udp: Utility-driven fetch directed instruction prefetching.
\newblock In {\em 2024 ACM/IEEE 51st Annual International Symposium on Computer
  Architecture (ISCA)}, pages 1188--1201. IEEE, 2024.

\bibitem{pagh2004cuckoo}
Rasmus Pagh and Flemming~Friche Rodler.
\newblock Cuckoo hashing.
\newblock {\em Journal of Algorithms}, 51(2):122--144, 2004.

\bibitem{bolt}
Maksim Panchenko, Rafael Auler, Bill Nell, and Guilherme Ottoni.
\newblock Bolt: a practical binary optimizer for data centers and beyond.
\newblock In {\em Proceedings of the 2019 IEEE/ACM International Symposium on
  Code Generation and Optimization}, CGO 2019, page 2–14. IEEE Press, 2019.

\bibitem{pellegrini2021arm}
Andrea Pellegrini.
\newblock Arm neoverse n2: Arm’s 2 nd generation high performance
  infrastructure cpus and system ips.
\newblock In {\em 2021 IEEE Hot Chips 33 Symposium (HCS)}, pages 1--27. IEEE,
  2021.

\bibitem{pellegrini2020arm}
Andrea Pellegrini, Nigel Stephens, Magnus Bruce, Yasuo Ishii, Joseph Pusdesris,
  Abhishek Raja, Chris Abernathy, Jinson Koppanalil, Tushar Ringe, Ashok
  Tummala, et~al.
\newblock The arm neoverse n1 platform: Building blocks for the next-gen
  cloud-to-edge infrastructure soc.
\newblock {\em IEEE Micro}, 40(2):53--62, 2020.

\bibitem{1238020}
E.~Perelman, G.~Hamerly, and B.~Calder.
\newblock Picking statistically valid and early simulation points.
\newblock In {\em 2003 12th International Conference on Parallel Architectures
  and Compilation Techniques}, pages 244--255, 2003.

\bibitem{pierce1996wrong}
Jim Pierce and Trevor Mudge.
\newblock Wrong-path instruction prefetching.
\newblock In {\em Proceedings of the 29th Annual IEEE/ACM International
  Symposium on Microarchitecture. MICRO 29}, pages 165--175. IEEE, 1996.

\bibitem{reinman1999fetch}
Glenn Reinman, Brad Calder, and Todd Austin.
\newblock Fetch directed instruction prefetching.
\newblock In {\em MICRO-32. Proceedings of the 32nd Annual ACM/IEEE
  International Symposium on Microarchitecture}, pages 16--27. IEEE, 1999.

\bibitem{rupley2018samsung}
J~Rupley.
\newblock Samsung exynos m3 processor.
\newblock {\em IEEE Hot Chips}, 30, 2018.

\bibitem{schall2022lukewarm}
David Schall, Artemiy Margaritov, Dmitrii Ustiugov, Andreas Sandberg, and Boris
  Grot.
\newblock Lukewarm serverless functions: characterization and optimization.
\newblock In {\em Proceedings of the 49th Annual International Symposium on
  Computer Architecture}, pages 757--770, 2022.

\bibitem{ignite}
David Schall, Andreas Sandberg, and Boris Grot.
\newblock Warming up a cold front-end with ignite.
\newblock In {\em Proceedings of the 56th Annual IEEE/ACM International
  Symposium on Microarchitecture}, MICRO '23, page 254–267, New York, NY,
  USA, 2023. Association for Computing Machinery.

\bibitem{srinivasan2001branch}
Viji Srinivasan, Edward~S Davidson, Gary~S Tyson, Mark~J Charney, and Thomas~R
  Puzak.
\newblock Branch history guided instruction prefetching.
\newblock In {\em Proceedings HPCA Seventh International Symposium on
  High-Performance Computer Architecture}, pages 291--300. IEEE, 2001.

\bibitem{stojkovicmosaic}
Jovan Stojkovic, Esha Choukse, Enrique Saurez, {\'I}nigo Goiri, and Josep
  Torrellas.
\newblock Mosaic: Harnessing the micro-architectural resources of servers in
  serverless environments.

\bibitem{qualcomm-oryon}
Gerard Williams.
\newblock { Qualcomm Oryon™ CPU }.
\newblock In {\em 2024 IEEE Hot Chips 36 Symposium (HCS)}, pages 1--21, Los
  Alamitos, CA, USA, August 2024. IEEE Computer Society.

\bibitem{wu2019temporal}
Hao Wu, Krishnendra Nathella, Joseph Pusdesris, Dam Sunwoo, Akanksha Jain, and
  Calvin Lin.
\newblock Temporal prefetching without the off-chip metadata.
\newblock In {\em Proceedings of the 52nd Annual IEEE/ACM International
  Symposium on Microarchitecture}, pages 996--1008, 2019.

\bibitem{recap}
Jason Zebchuk, Harold~W. Cain, Xin Tong, Vijayalakshmi Srinivasan, and Andreas
  Moshovos.
\newblock Recap: A region-based cure for the common cold (cache).
\newblock In {\em 2013 IEEE 19th International Symposium on High Performance
  Computer Architecture (HPCA)}, pages 83--94, 2013.

\bibitem{zhang2025hierarchical}
Tingji Zhang, Boris Grot, Wenjian He, Yashuai Lv, Peng Qu, Fang Su, Wenxin
  Wang, Guowei Zhang, Xuefeng Zhang, and Youhui Zhang.
\newblock Hierarchical prefetching: A software-hardware instruction prefetcher
  for server applications.
\newblock In {\em Proceedings of the 30th ACM International Conference on
  Architectural Support for Programming Languages and Operating Systems, Volume
  2}, pages 529--544, 2025.

\bibitem{ocolos}
Yuxuan Zhang, Tanvir~Ahmed Khan, Gilles Pokam, Baris Kasikci, Heiner Litz, and
  Joseph Devietti.
\newblock Ocolos: Online code layout optimizations.
\newblock In {\em Proceedings of the 55th Annual IEEE/ACM International
  Symposium on Microarchitecture}, MICRO '22, page 530–545. IEEE Press, 2023.

\end{thebibliography}

\end{document}